\documentclass[useAMS,usenatbib]{mn2e} 
\usepackage{aas_macros}
\usepackage{graphics}
\usepackage[pdftex]{graphicx}
\usepackage{epstopdf}
\usepackage{epsfig}  
\usepackage{natbib} 
\usepackage{float}
\usepackage{amsmath}
\usepackage{times}
\usepackage[varg]{txfonts}
\usepackage{verbatim} 
\usepackage{multirow,bigdelim} 
\usepackage{color}
\usepackage{vmargin}

\setmarginsrb{1.2cm}{1cm}{1.2cm}{1cm}{1cm}{1cm}{1cm}{1cm}

\usepackage{multirow}
\usepackage{amssymb}
\usepackage{color}
\usepackage{boxedminipage}
\usepackage{rotating}
\usepackage{ulem}

\def \kms {{\rm km s$^{-1}$}}
\def \kmsMpc {{\rm km s$^{-1}$ Mpc$^{-1}$}}

\newcommand{\archangel}{{\sc archangel}}

\title[Photometry and TF distances]{From Spitzer Galaxy Photometry to Tully-Fisher Distances}
\author[Sorce et al.]
{J. G. Sorce$^{1,2}$\thanks{E-mail: \texttt{j.sorce@ipnl.in2p3.fr}}, 
R. B. Tully$^{3}$,
H. M. Courtois$^{1}$,
T. H. Jarrett$^{4}$,
J. D. Neill$^{5}$,
E. J. Shaya$^{6}$\\
$^1$Universit\'e Lyon 1, CNRS/IN2P3, Institut de Physique Nucl\'eaire, Lyon, France\\
$^2$Leibniz-Institut f\"{u}r Astrophysik, Potsdam, Germany\\
$^3$Institute for Astronomy, University of Hawaii, 2680 Woodlawn Drive, HI 96822, USA\\
$^4$University of Cape Town, Private Bag X3, Rondebosch 7701, Republic of South Africa\\
$^5$California Institute of Technology, 1200 E. California Blvd. MC 278-17, Pasadena, CA 91125, USA\\
$^6$Department of Astronomy, University of Maryland, College Park, MD 20742, USA\\
}
\begin{document}

\date{}

\maketitle

\label{firstpage}

\begin{abstract}
This paper involves a data release of the observational campaign: {\it Cosmicflows with Spitzer} (CFS). Surface photometry of the 1270 galaxies constituting the survey is presented. An additional $\sim$ 400 galaxies from various other Spitzer surveys are also analyzed. CFS complements the {\it Spitzer Survey of Stellar Structure in Galaxies}, that provides photometry for an additional 2352 galaxies, by extending observations to low galactic latitudes ($|b|<30^\circ$). Among these galaxies are calibrators, selected in K band, of the Tully-Fisher relation. The addition of new calibrators demonstrate the robustness of the previously released calibration. Our estimate of the Hubble constant using supernova host galaxies is unchanged, H$_0 = 75.2 \pm3.3$~\kmsMpc. Distance-derived radial peculiar velocities, for the 1935 galaxies with all the available parameters, will be incorporated into a new data release of the Cosmicflows project. The size of the previous catalog will be increased by 20\%, including spatial regions close to the Zone of Avoidance.
\end{abstract}



\section{Introduction}

{\it Cosmicflows} \citep{2012ApJ...749...78T,2012ApJ...749..174C,2012AN....333..436C,2013AJ....146...86T} is a project to map radial peculiar velocities of galaxies within 200 Mpc with the ultimate goal of reconstructing and simulating the motions of the large-scale structures and explaining the deviation of our Galaxy from the Hubble expansion of 630 \kms\ \citep{1996ApJ...473..576F}. Radial peculiar velocities, v$_{pec}$, are obtained from the redshift and an independent luminosity distance measurement, v$_{pec}$ = v$_{mod}$ - H$_0$ d, where H$_0$ is the Hubble constant and v$_{mod}$ is the velocity with respect to the Cosmic Microwave Background with a minor correction for cosmological effects \citep[][]{2013AJ....146...86T}. Distances in the project {\it Cosmicflows} are mainly obtained with the luminosity-linewidth rotation rate correlation or Tully-Fisher relation \citep[TFR,][]{1977A&A....54..661T}, a distance estimator which provides coverage up to 200 Mpc. The TFR necessitates two very accurate observations of a galaxy to compute its distance - an HI profile and a photometric measurement. Observations in the radio domain to obtain rotation rates of galaxies have made great advances in the past few years and more than 10,000 adequate linewidths of galaxies are available \citep{2011MNRAS.414.2005C} in the Extragalactic Distance Database\footnote{http://edd.ifa.hawaii.edu/} \citep[EDD,][]{2009AJ....138..323T,2009AJ....138.1938C}. The {\it Cosmicflows with Spitzer} (CFS) program, combined with an additional sample of galaxies from various Spitzer programs, uses the space-based Spitzer telescope \citep{2004ApJS..154....1W} to address the photometric requirement of the project {\it Cosmicflows}.\\

In this paper, we present the reduction of wide-field images of 1270 galaxies observed with the 3.6 $\mu$m channel of the InfraRed Array Camera \citep[IRAC,][]{2004ApJS..154...10F} onboard the Spitzer space telescope during its post-cryogenic period, cycle 8. This survey complements four other large Spitzer surveys, the Spitzer Infrared Nearby Galaxy Survey \citep[SINGS,][]{2009ApJ...703.1569M}, the Local Volume Legacy Survey \citep[LVL,][]{2009ApJ...703..517D} led during the cryogenic phase of the Spitzer mission, the Carnegie Hubble Program \citep[CHP,][]{2011AJ....142..192F} and the Spitzer Survey of Stellar Structure in Galaxies \citep[S$^4$G,][]{2010PASP..122.1397S} obtained in the post-cryogenic period. From these surveys and several other small programs, approximately 1000 additional galaxies are of interest to the {\it Cosmicflows} project. Approximately 35\% of these galaxies are reduced using the Spitzer-adapted version of \archangel\ while S$^4$G-pipeline (Mun\~oz-Mateos et al. in prep.) supplies the rest of it. With the availability of such a large number of photometric measurements, the robustness of both the TFR calibration method and the TFR at 3.6 microns can be confirmed. 

In the subsequent section, we describe the complete photometric sample, then we present the observation-reduction process applied to CFS and approximately 400 supplementary galaxies and the results. In the third section, the
mid-IR TFR \citep[although at that time considered preliminary]{2013ApJ...765...94S} is shown to be robust. The associated Hubble constant estimate is confirmed in the fourth section. In the last section, we derive accurate distance estimates for 1935 galaxies with acceptable inclinations and available linewidths that either we have reduced or that come from the S$^4$G analysis.


\section{Observational samples}
\label{observation}

\begin{figure}
\centering
\includegraphics[scale=0.5]{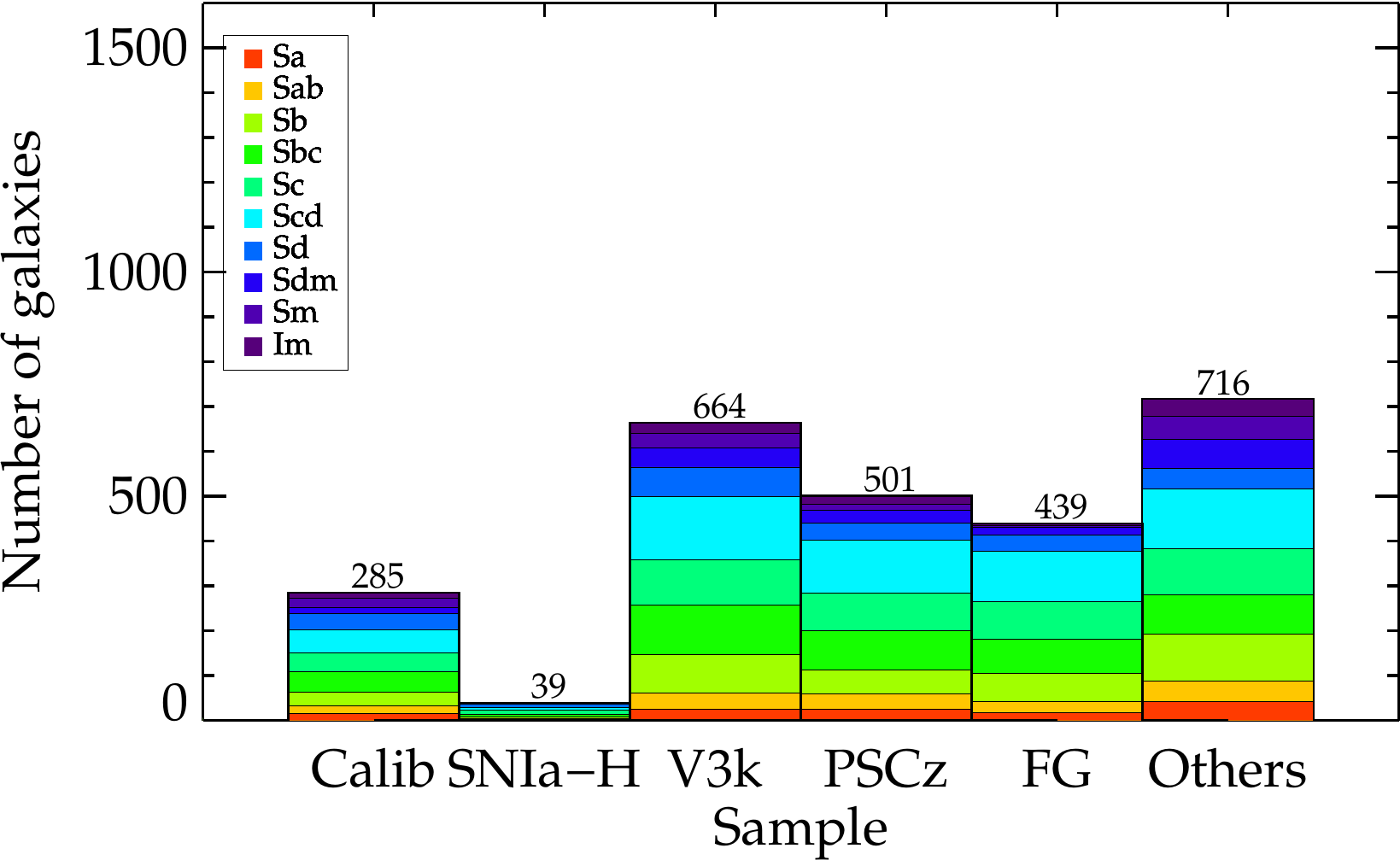}
\caption{Histogram of the number of galaxies per subsamples in CFS and diverse programs, mostly S$^4$G (65\%). Calib is constituted of TFR calibrators, SNIa-H contains hosts of SNIa, V3k is built of galaxies with v$_{hel}$ $<$ 3000 \kms\ , PSCz is derived from the IRAS point-source redshift survey and FG is a catalog of flat galaxies. "Others" stands for galaxies of interests which do not fall into one of the previously cited categories. The gradient of colors shows the proportion of each morphological type from the HyperLeda Database in each sample.} 
\label{sample}
\end{figure}

In Figure \ref{sample}, the 1270 galaxies of the CFS survey are distinguished by their occurrence in five subsamples: 1. the TF calibrators (Calib), 2. the hosts of SNIa sample (SNIa-H), 3. the V3k, 3000 \kms\ sample (V3k), 4. the IRAS point source-redshift sample (PSCz) and 5. the flat galaxy sample (FG). These subsamples are completed with  galaxies from various surveys. If a galaxy lies within multiple samples, in the following the galaxy is assigned to the sample that includes it that is discussed first. Galaxies of interest to the project but which do not fall into one of the previous categories constitute the sixth subsample. All these supplementary galaxies are mostly from S$^4$G (65\%). Among the $\sim$ 400 galaxies left, most galaxies have been observed by SINGS (2\%), LVL (3\%) and CHP (16\%) programs. 

\begin{figure}
\centering
\includegraphics[scale=0.615]{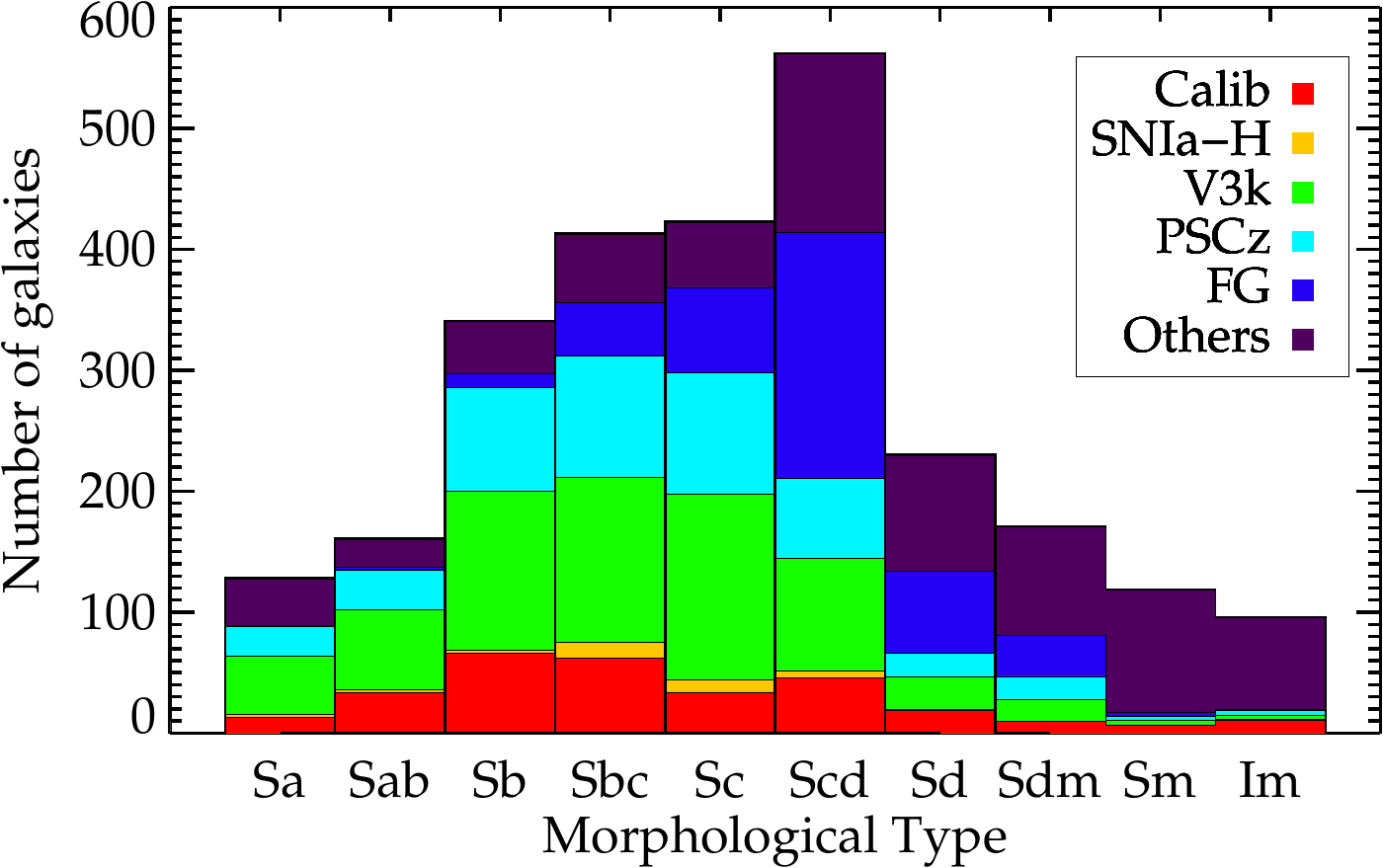}
\includegraphics[scale=0.6]{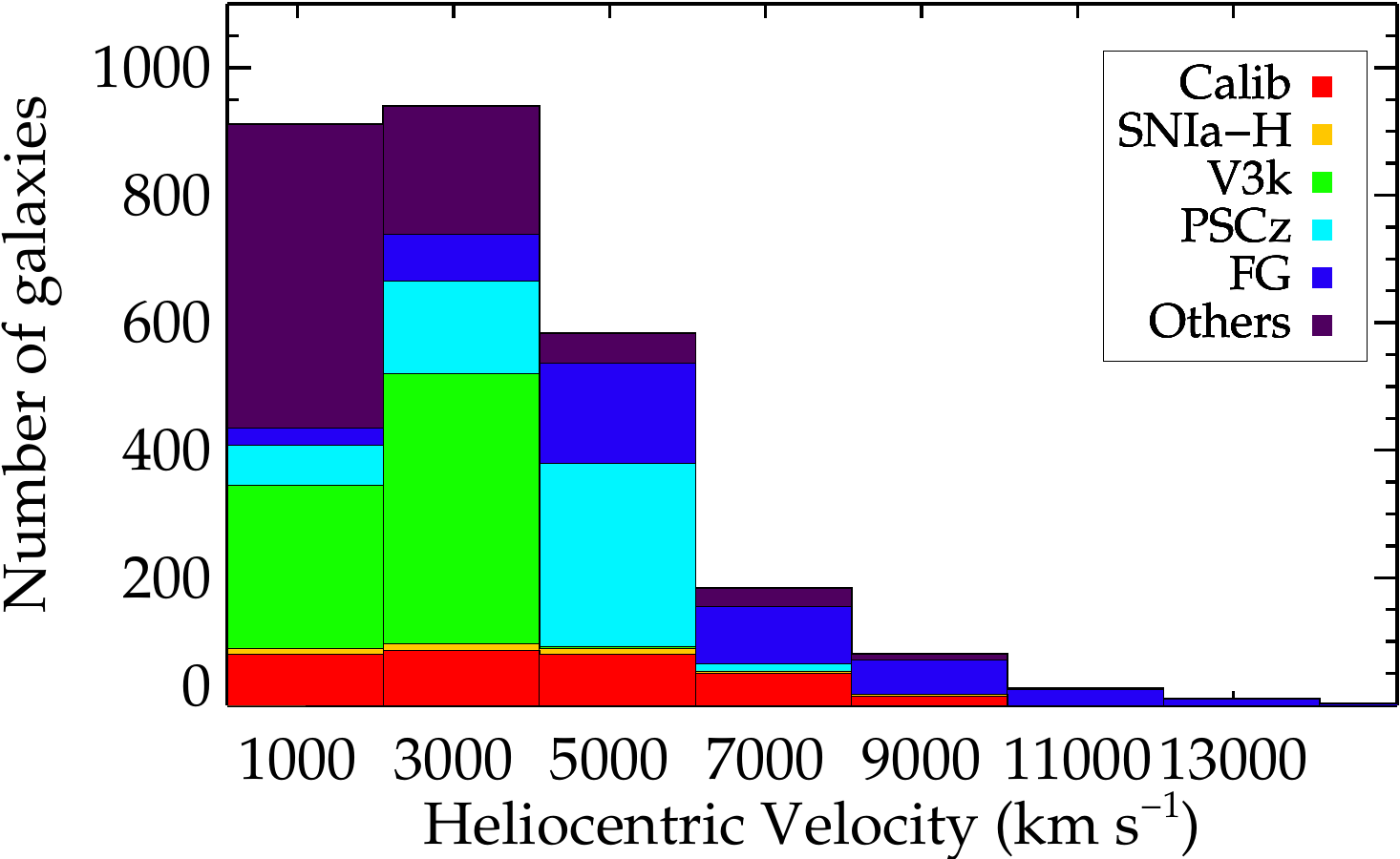}
\caption{Histograms of the morphological type (top) from HyperLeda and of the heliocentric velocity (bottom) from EDD for the whole compilation of galaxies. The gradient of colors gives in which proportion each subsample contribute to a given type (top) and range of heliocentric velocities (bottom).}
\label{type-vhel}
\end{figure}

\begin{figure*}
\vspace{-2cm}
\centering
\includegraphics[scale=0.55]{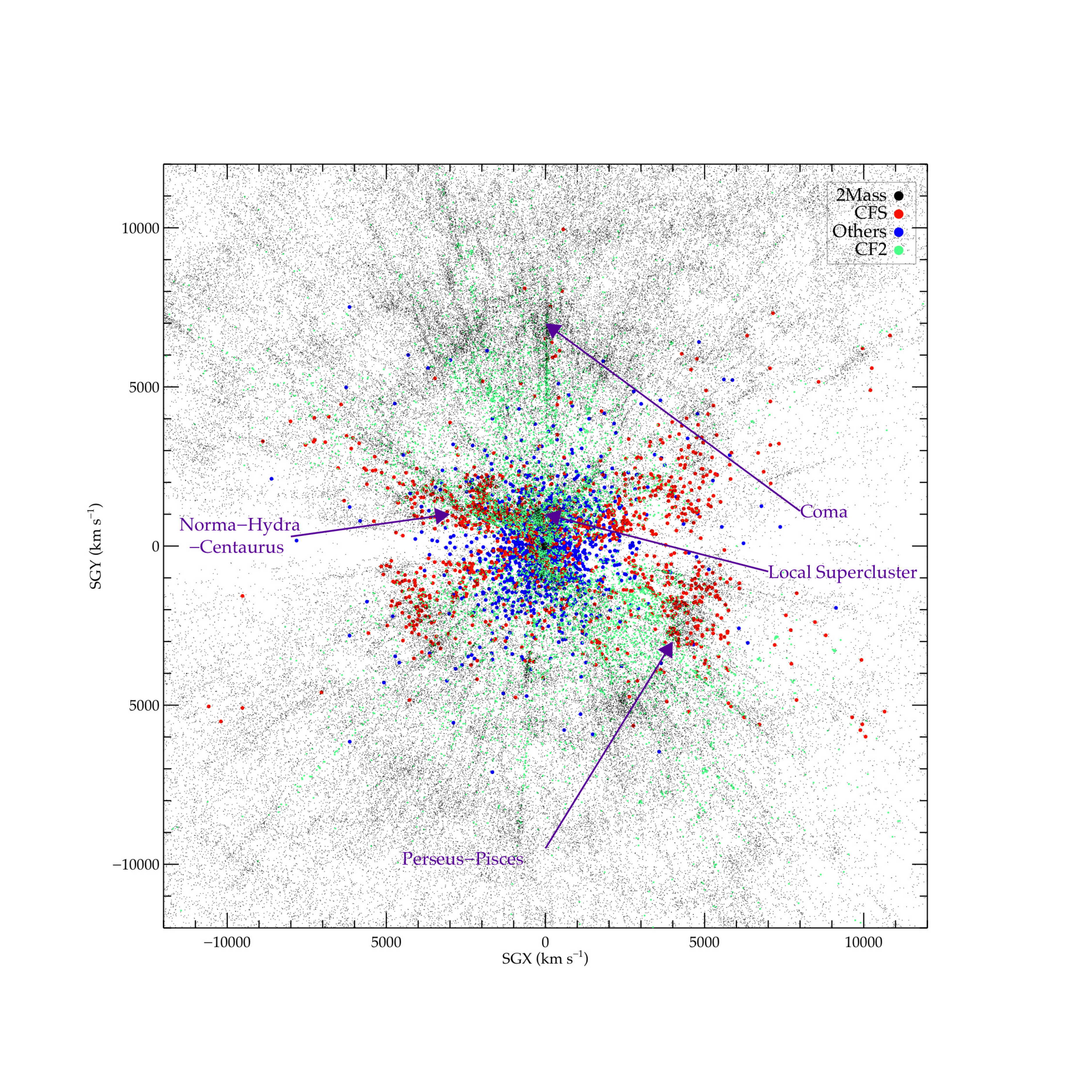}
\vspace{-1.cm}
\caption{In the XY supergalactic plane, galaxies of the CFS survey (red dots) are superimposed on the 2MASS redshift catalog (tiny black dots). Blue dots stand for galaxies of interests to the {\it Cosmicflows} project but observed by different programs, mostly S$^4$G. A few superclusters are identified by violet arrows. CFS completes previous surveys with galaxies at low galactic latitudes. Green dots represents the second catalog of the Cosmicflows project. Future catalogs of the Cosmicflows project will have a better coverage near the Zone Of Avoidance, reconstructions of the Local Universe will be more accurate in that region.}
\label{position}
\end{figure*}
      
- The first two of these subsamples have already been described \citep{2000ApJ...533..744T,2012ApJ...749...78T,2012ApJ...749..174C} and partly used at 3.6 $\mu m$ to calibrate the TFR in the mid-infrared \citep{2013ApJ...765...94S} and to define an absolute zero-point to the SNIa scale \citep{2012ApJ...758L..12S} respectively. Approximately one third of the first subsample is constituted of galaxies observed for CFS. Others have been observed by previous Spitzer programs, mostly CHP and S$^4$G. Half of the SNIa subsample is made of CFS observations while the other half contains mostly CHP observations. \\ 
- The third subsample is a catalog developed over the years called V3k \citep{2008ApJ...676..184T}. It extends up to the velocity limit, 3000 \kms\ , imposed by the capabilities of early-generation radio telescopes to obtain useable HI profiles and gives coverage of the traditional Local Supercluster \citep{1953AJ.....58...30D}. Figures \ref{sample} and the top of Figure \ref{type-vhel} show that the majority of these galaxies are of types later than Sa. Types come from the HyperLeda database \citep{2003A&A...412...45P}. Figure \ref{type-vhel} bottom confirms that the heliocentric velocities of these galaxies are $V_h <$ 3300 \kms\ with heliocentric velocities coming from EDD. Among the 683 galaxies available for this third subsample about a quarter comes from the CFS survey. This sample provides a high density of the Local Supercluster centered on Virgo. \\
- The next subsample is based on the redshift survey PSCz \citep{2000MNRAS.317...55S} of sources drawn from a flux-limited sample at $100 \mu$m obtained with the InfraRed Astronomical Satellite. 
The sample is dominated by normal spirals distributed around the Sc type as Figures \ref{sample} and \ref{type-vhel} show. The heliocentric velocity limit is 6000 \kms\ to obtain reasonable HI lines with current radio telescopes. This subsample includes the Norma-Hydra-Centaurus and the Perseus-Pisces superclusters in the opposite directions and many low latitude galaxies - offering good coverage above $|b| = 5^\circ$. The bifurcation between our flow direction and a motion towards Perseus-Pisces highlighted by \cite{2006MNRAS.373...45E} will be located thanks to this subsample. The PSCz sample will also strongly constrain the CMB dipole component within 6000 \kms. CFS contains the majority (445) of these galaxies.\\
- The last subsample is constituted of flat galaxies from the catalog of \citet{1999BSAO...47....5K}. These edge-on systems have a major to minor axis ratio greater than 7 implying minimal de-projection of their HI linewidths. The flat galaxies are principally of type Scd, as shown in Figures \ref{sample} and \ref{type-vhel} top. They constitute a homogeneous class of HI rich systems but they have a low space density partly because of  the strong inclination constraint. Extinction problems existing at optical bands and for ground-based telescoped are practically removed with IRAC 3.6 microns. The entire flat galaxy subsample comes from CFS observations.\\ 

Figure 3 illustrates the combined coverage of CFS and other relevant surveys with Spitzer Space Telescope. CFS gives special attention to galaxies at low galactic latitudes for two reasons.  First, CFS complements the important S$^4$G survey that has a $|b|=30^\circ$ lower limit and that supplies most of the other galaxies.  Second, we recognize that photometry from WISE, the Wide-Field Infrared Survey Explorer \citep{2010AJ....140.1868W}, will be useful but be at a competitive disadvantage to Spitzer in the crowded star fields at lower galactic latitudes because of resolution issues. As a result, future catalogs of the Cosmicflows project will contain more data close to the Zone Of Avoidance than the second catalog (CF2) of the project superimposed on the same figure.\\

In section \ref{reduction}, a comparison between 241 magnitudes from S$^4$G-pipeline and from the Spitzer-adapted version of \archangel\ used in this paper reveals the very good agreement between both magnitudes. As a result, S$^4$G-magnitudes are directly used to derive distances for the relevant galaxies in the last section. In the next section, we focus mostly on the CFS sample although the additional Spitzer archival galaxies minus S$^4$G's are processed equally.


\section{Reductions, Analyses and Comparisons}
\label{reduction}

\subsection{Reductions}
The Post-Basic Calibrated Data of the 1270 observed galaxies for the CFS program are available at the Spitzer Heritage Archive. Every galaxy has been observed with the first channel of the IRAC instrument where a point spread function with a FWHM 1.66" is sampled with 1.2" pixels. The field of view is 5.2 x 5.2 arcmins which is adequate to include most galaxies beyond twice their diameter at the 25$^{th}$ isophote (mag arcsec$^{-2}$) in B band. Consequently, except for a few cases, galaxies (1219 out of 1270) have been mapped within a single field exposed during four minutes (the total duration of one observation is 8.6 min), 45 have been mapped with four fields, five with nine fields and one, PGC62836 (NGC6744), with 16 fields. Every resulting composite field extends to 8.5 exponential scale lengths ensuring that 99\% of the light of the galaxy is captured. The galaxies have inclination  i $>$ 45$^\circ$ and are not perturbed by - or confused with - a second object in the HI beam ensuring that the TFR can be applied to them later on with minimized uncertainties. \\

The photometry is carried out with a Spitzer-adapted version of \archangel\ \citep{2007astro.ph..3646S,2012PASA...29..174S} described in detail in \citet{2012AJ....144..133S}. Briefly, \archangel\ performs the masking of stars and flaws and it replaces masked regions by mean isophote values. It fits ellipses to isophotes with increasing radii. It compresses the 2D information into unidimensional surface brightness and magnitude growth curves. Finally, parameters such as extrapolated magnitudes are derived. We run \archangel\ twice on each galaxy. The first run supplies the second run with parameters to improve the results. In the second run, we force the ellipse fitting up to at least 1.5 $\times$ a$_{26.5}$ - radius of the 26.5 mag arcsec$^{-2}$ isophote at 3.6 $\mu$m - to ensure that 99\% of a galaxy light is captured. Position angles and ellipticities are frozen only at large radii - basically a$_{24}$, radius of the 24 mag arcsec$^{-2}$ isophote at 3.6 $\mu$m  - where the noise dominates, except when a simple vizualisation shows that a smaller freezing radius is required. Very flat galaxies are mostly among the exceptions where the masking fails without a smaller freezing radius: the edges are inevitably masked if ellipses are not frozen at small radii. In any case, position angles and ellipticities at medium radii overall do not affect magnitudes, the most important parameter for the TFR. \citet{2012AJ....144..133S} showed that the major contribution to the magnitude uncertainties is the sky setting. Every source of uncertainty included, the total magnitude uncertainty is still held below 0.04 mag (0.05 with extinction, aperture and k-corrections) for normal spiral galaxies.\\

For each galaxy, we derive the major axis radius in arcsec of the isophote at 26.5 mag arcsec$^{-2}$ in the [3.6] band, a$_{26.5}$, and of the annuli enclosing respectively 80\%, 50\% and 20\% of the total light, a$_{80}$, a$_e$ and a$_{20}$ (the subscript ``e'' stands for ``effective'' - a common terminology). We compute also the corresponding surface brightnesses $\mu_{80}$, $\mu_e$ and the average $<\mu_{e}>$ of the surface brightnesses between 0 and 50\% of the light, $\mu_{20}$ and the average $<\mu_{20}>$ between 0 and 20\% of the light, in mag arcsec$^{-2}$. The central disk surface brightness in mag arcsec$^{-2}$, $\mu_0$, the exponential disk scale length in arcsec, $\alpha$, the mean b/a ratio and its variance, the position angle and the concentration index a$_{80}$/a$_{20}$ are also given. Three magnitudes are calculated: the magnitude at the 26.5$^{th}$ isophote, [3.6]$_{26.5}$, the total magnitude obtained from the extrapolation of the growth curve, [3.6]$_{tot}$ (the uncertainty on the rational function fit used to derive [3.6]$_{tot}$ is also given) and the extrapolated magnitude assuming a continuous exponential disk, [3.6]$_{ext}$. All the magnitudes are given in the AB system. We recommend to use [3.6]$_{ext}$ even if the three magnitudes are very similar.\\

Isophotes, surface brightness profiles and growthcurves are available for the 1270 galaxies on line along with a table of the derived parameters at the EDD website. These plots are also available in EDD for the additional $\sim$ 400 galaxies from other programs drawn from the Spitzer archive.

\subsection{Analyses}

\begin{figure}
\centering
\includegraphics[scale=0.45]{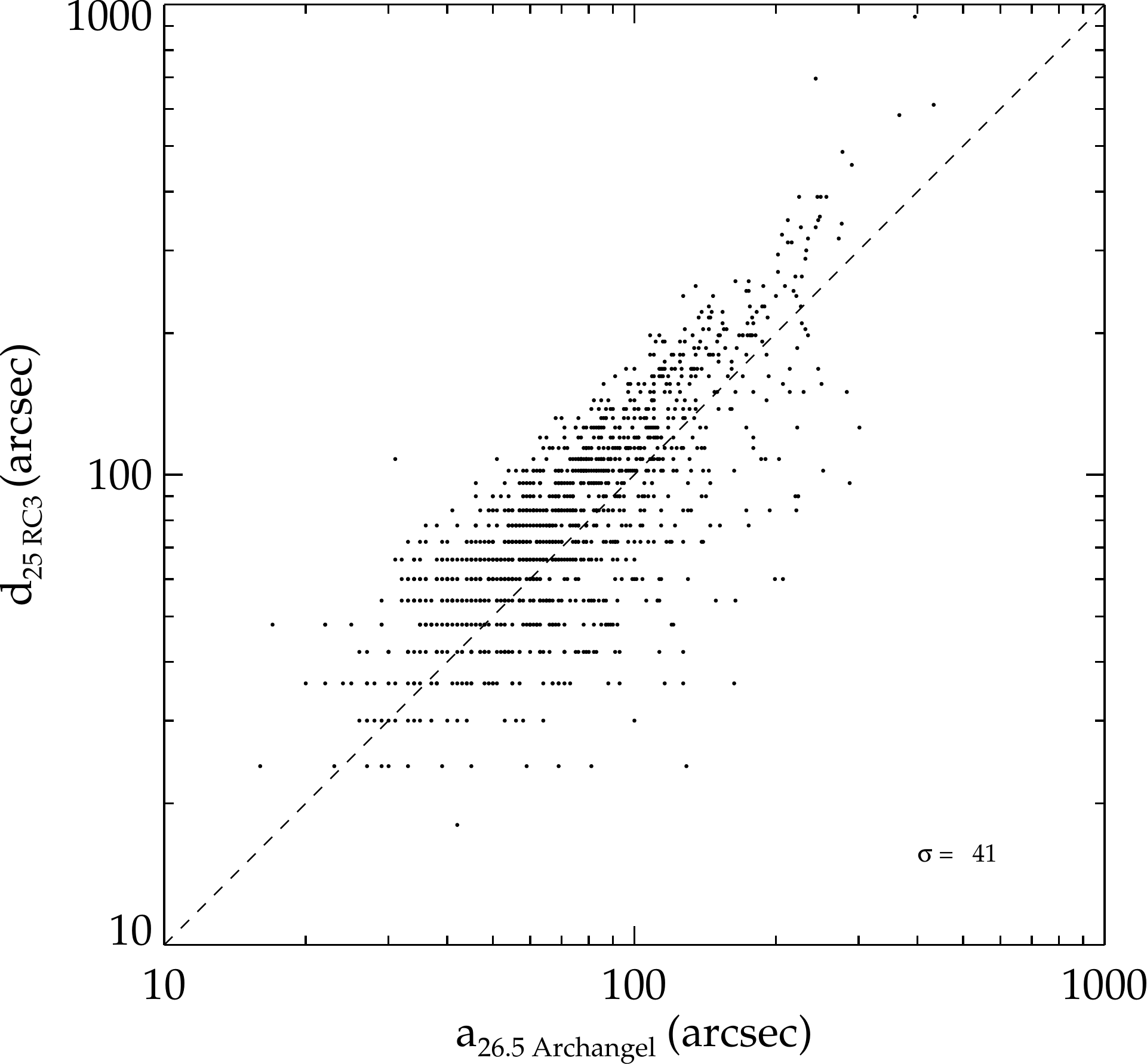}
\caption{Comparison between the radius in arcsec of the isophote at 26.5 mag arcsec$^{-2}$ in the [3.6] band obtained after reduction with \archangel\ and the radius at 25 mag arcsec$^{-2}$ at B band used beforehand to set observational parameters. These parameters are proportional to each other. In the case of an optimal 1:1 linear relation, the scatter is only 41 arcsec.}
\label{comp}
\end{figure}

\begin{figure}
\centering
\includegraphics[scale=0.6]{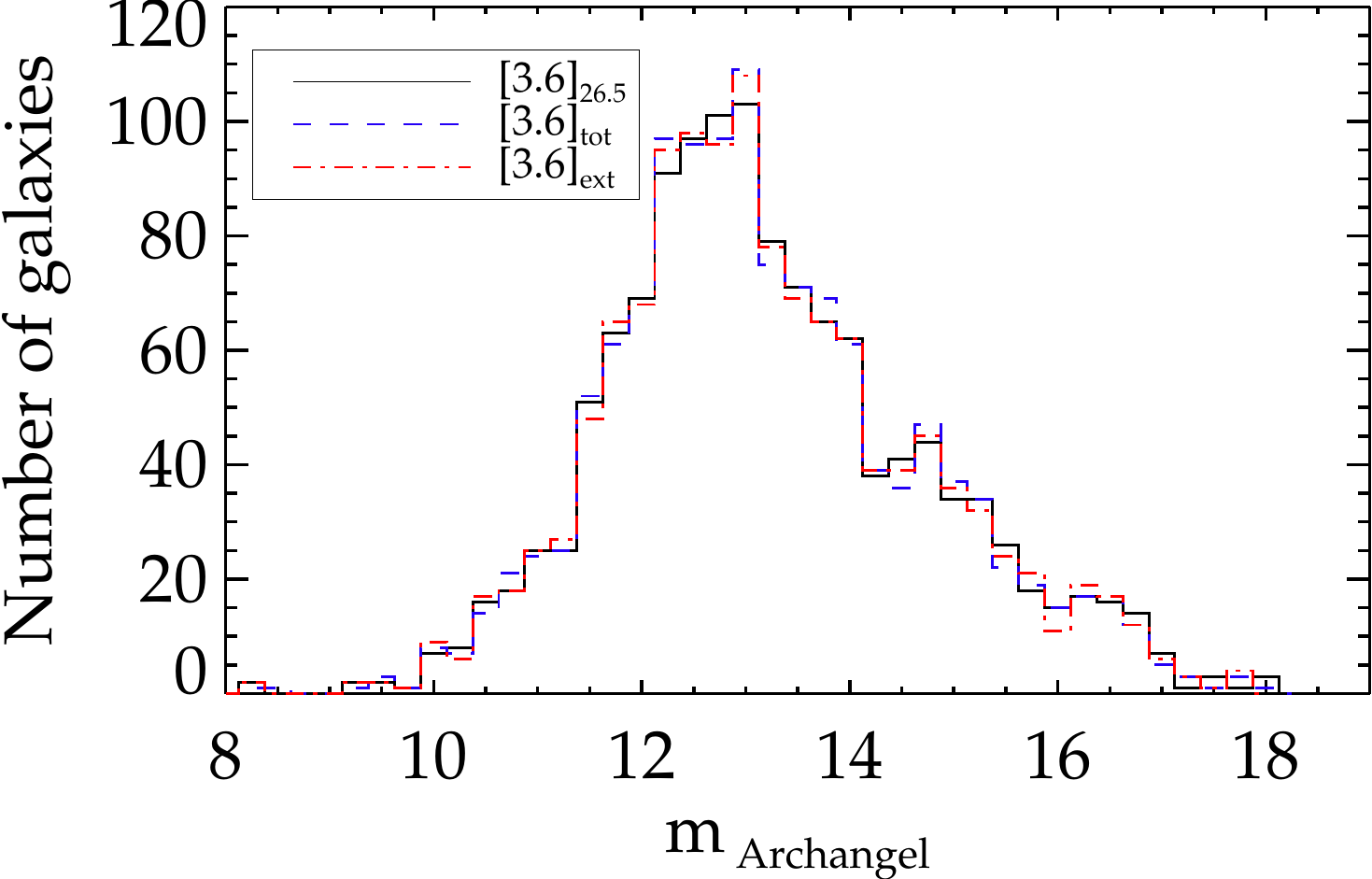}
\caption{Histograms of the three magnitudes derived with \archangel\. The magnitude at the 26.5 mag arcsec$^{-2}$ isophote at 3.6 $\mu m$, [3.6]$_{26.5}$ (black straight line), the magnitude obtained by the extrapolation of the growth curve, [3.6]$_{tot}$ (blue dashed line) and the magnitude assuming a continuous exponential disk, [3.6]$_{ext}$ (red dotted-dashed line).}
\label{mag}
\end{figure}

\begin{figure}
\centering
\includegraphics[scale=0.525]{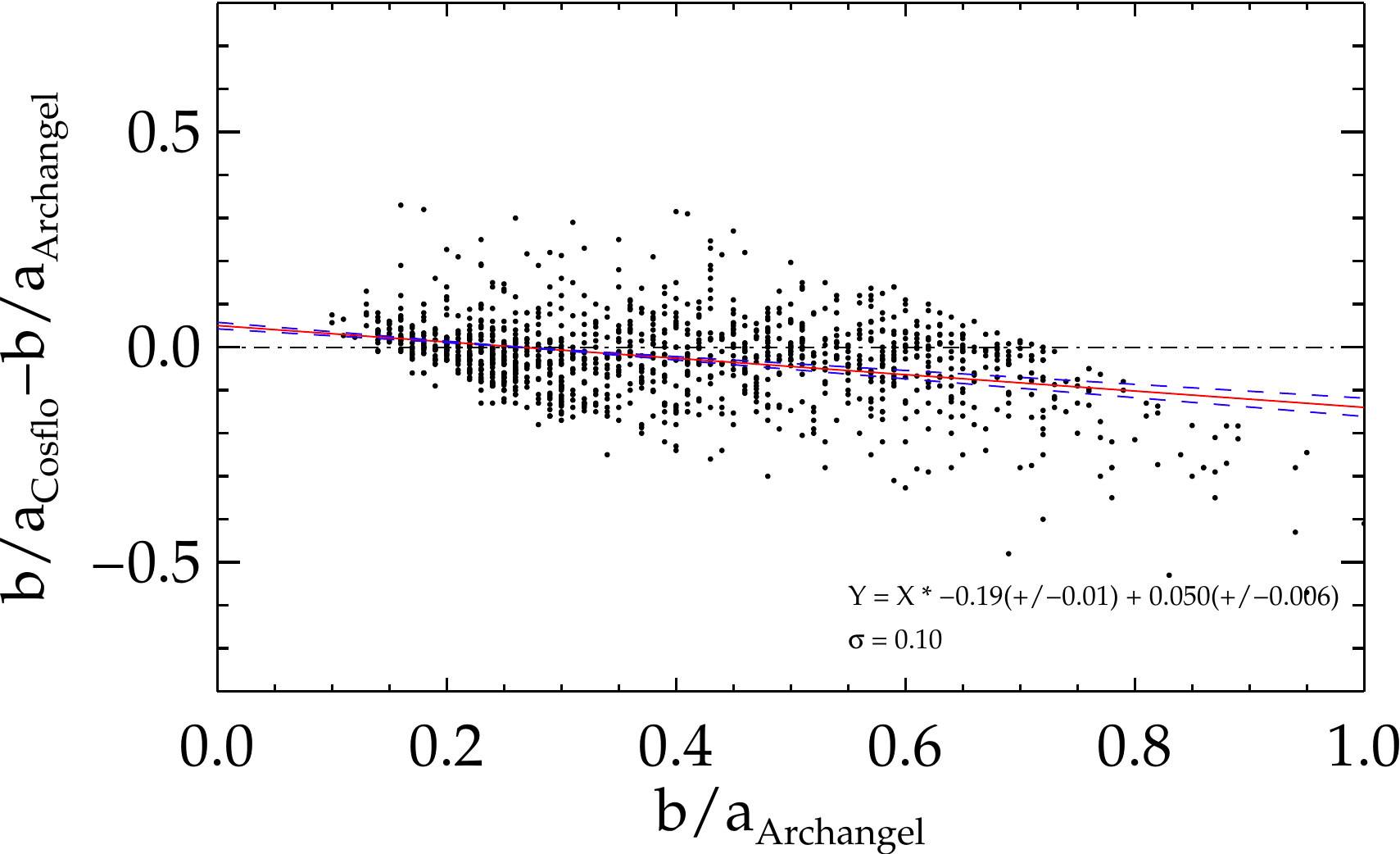}
\includegraphics[scale=0.53]{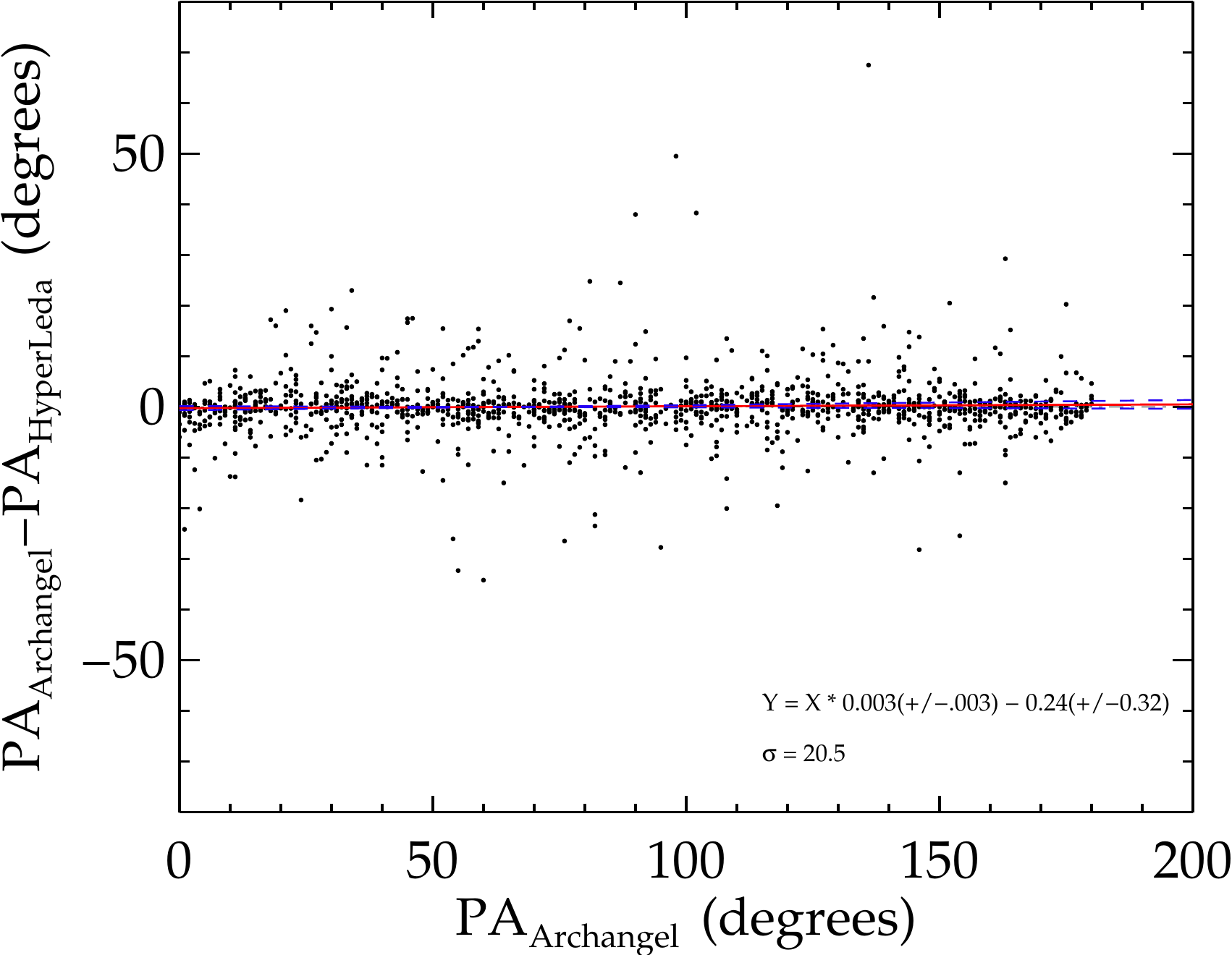}
\caption{\archangel-derived b/a ratios (top) and position angles (bottom) vs Cosmicflows' minus \archangel's and Hyperleda's minus \archangel's. The black dotted-dashed lines show the perfect cases y=0, the red straight lines are linear fits to the data (with a 3-$\sigma$ clipping (20 galaxies) in the bottom panel), the blue dashed lines are the 1-$\sigma$ uncertainties.}
\label{ba-pa}
\end{figure}

In this subsection, we present the different parameters derived with the software \archangel\ for each one of the CFS galaxies. We claim at the beginning of section \ref{reduction} that we choose to observe each galaxy to within at least twice d$_{25}$ to capture most of galaxy lights and to minimize magnitude measurement uncertainties. Then, we force ellipse fitting up to 1.5 $\times$ a$_{26.5}$. Figure \ref{comp} confirms that d$_{25}$ from RC3 used to set observations and a$_{26.5}$ obtained after reduction are comparable representatives of size. The scatter is only 41 arcsec around a 1:1 linear relation. The observational sensitivity is sufficient for our ultimate goal since at 26.5 mag arcsec$^{-2}$ the isophotal magnitudes are already very close to extrapolated ones as shown by \citet{2012AJ....144..133S}  and Figure \ref{mag}.\\

In the adapted version of \archangel\, the computation of the minor to major axis, b/a, ratio, is specifically defined as the mean of the b/a ratios between 50 and 80\% of the light. Measuring b/a ratios is not an easy task and a comparison with the ratios used in the Cosmicflows program on Figure \ref{ba-pa} top shows that at least one b/a source cannot be trusted. Each value needs to be checked before any usage. Retained b/a values are from the I band program of Cosmicflows \citep{2013AJ....146...86T} and from HyperLeda if it comes from \citet{2003A&A...412...45P}. Position angles on the other hand are in good agreements at the bottom of the same Figure.

\begin{figure*}
\centering
\includegraphics[scale=1]{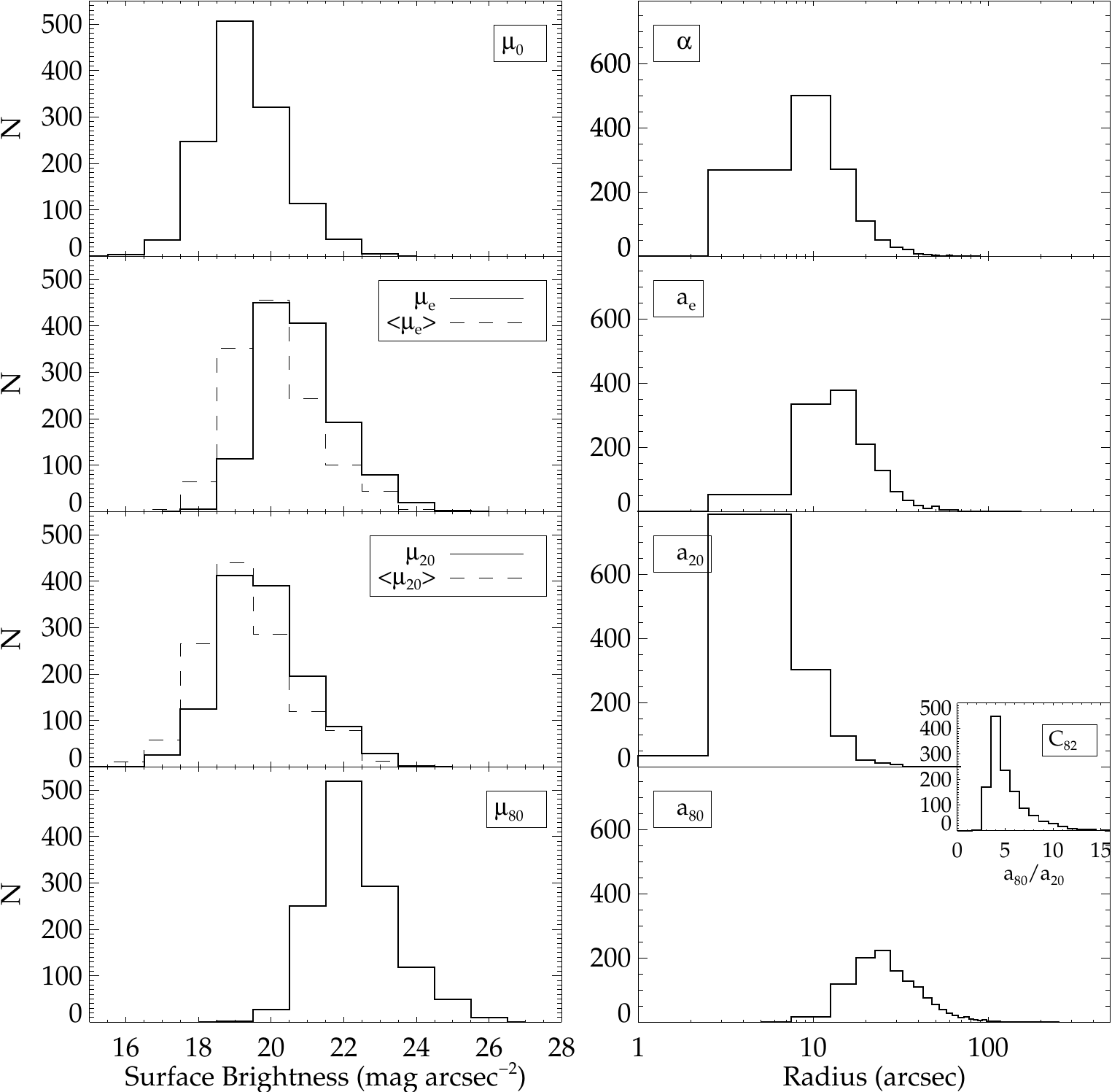}
\caption{Histograms of some of the parameters computed with the \archangel\ software. Left, from top to bottom, histograms in solid lines of the central disk surface brightness $\mu_{0}$ and of the surface brightnesses at 50, 20 and 80\% of the total light $\mu_e$, $\mu_{20}$ and $\mu_{80}$ (mag arcsec$^{-2}$). Histograms of the average of the surface brightnesses between 0 and 50 and 20\% of the light, $<\mu_{e}>$ and $<\mu_{20}>$  respectively, are overplotted in dashed lines. Right, from top to bottom, disk scale length $\alpha$ and annuli encompassing 50, 20 and 80\% of the light a$_e$, a$_{20}$ and a$_{80}$, in arcsec. The histogram of the concentration index, C$_{82}$ = a$_{80}$ / a$_{20}$ is overplotted in a small panel on the right side of the a$_{20}$ and a$_{80}$ histograms.}
\label{all}
\end{figure*}

Histograms of the other parameters are given in Figure \ref{all} in mag arcsec$^{-2}$ for surface brightnesses and in arcsec for corresponding radii. For all these parameters there is no outliers. It is worth noting that [3.6] micron surface brightnesses are overall below 24 mag arcsec$^{-2}$ which is better than most optical surveys (about 26-28 mag arcsec$^{-2}$ in B-band for example). IRAC is an exquisite imager for flatness and depth.\\

\subsection{Comparisons}

\begin{figure}
\centering
\includegraphics[scale=0.55]{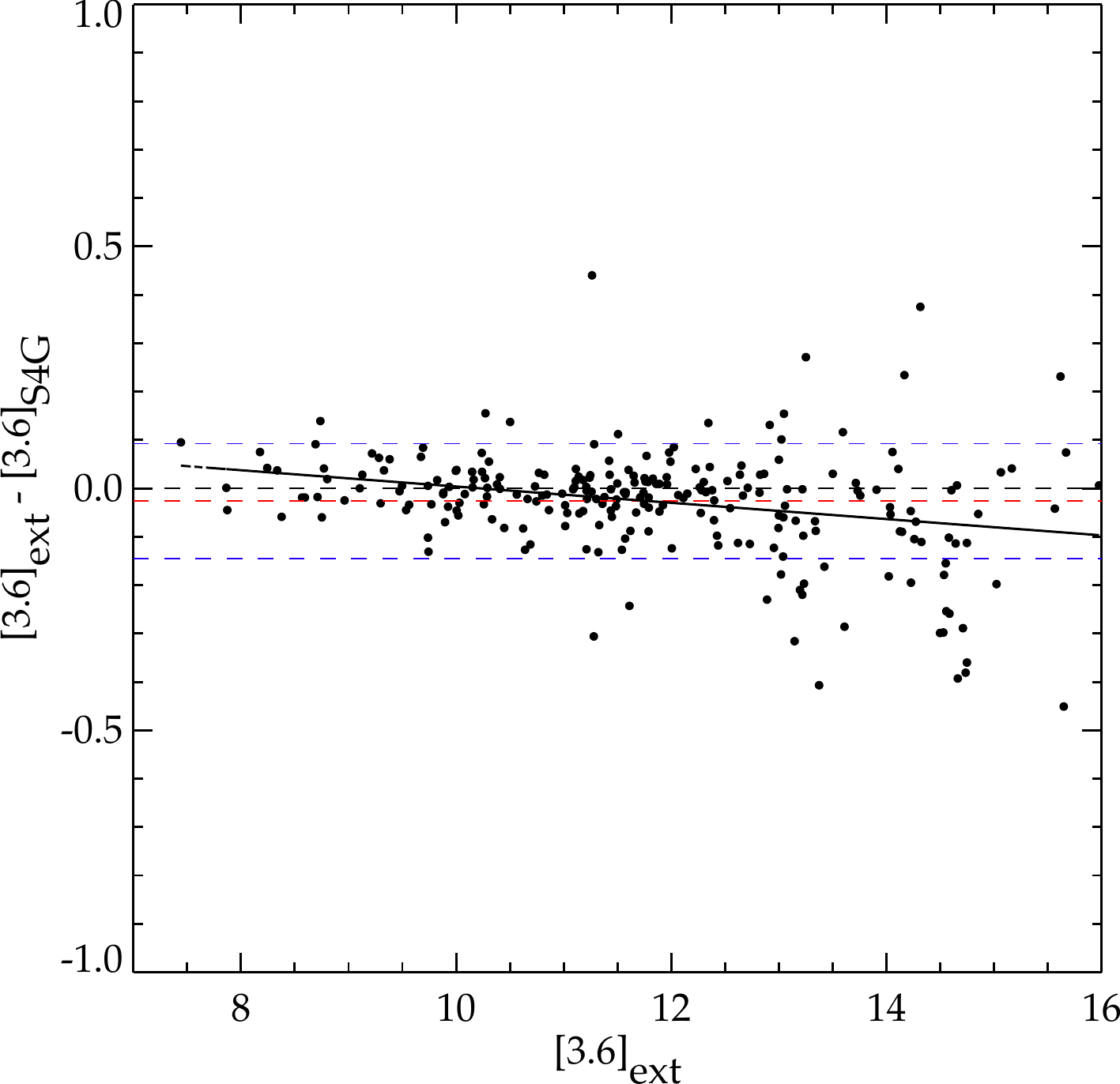}
\caption{Comparisons between 241 [3.6] extrapolated \archangel\ and [3.6] S$^4$G magnitudes. The fit at 3-$\sigma$ clipping (4 galaxies rejected) has a slope of -0.02 $\pm$ 0.004 and a zero point of 0.17 $\pm$0.04. The red dashed thick line stands for the offset at -0.02 mag and the blue dashed lines represent the scatter at 0.1 mag. Deviant cases except for 2 are low surface brightness galaxies and we find no reason to reject \archangel\ values.}
\label{S4G}
\end{figure}

This last subsection demonstrates the agreement between magnitudes obtained with the Spitzer-adapted version of \archangel\ used in this paper and with alternative pipelines. Figure 9 of \citet{2012AJ....144..133S} had already revealed that \archangel\ and the software developed for the GALEX Large Galaxy Atlas (GLGA, Seibert et al., in prep.) by the CHP team give relatively close magnitudes. Figure \ref{S4G} proves that this adapted version of \archangel\ computes magnitudes equally similar to the pipeline of the S$^4$G team. 
There is a slight tendency for S$^4$G values to be brighter for the largest galaxies. A cause can be the difference in masking. Another cause can be the sky setting that with S4G is done quite differently. Instead of using sky boxes, S$^4$G pipeline derives sky values out of annuli located just at the extremity of what they estimate to contain the totality of the galaxy light. This different sky setting might also explain the slight increase in the root mean square scatter (4 galaxies rejected) which reaches $\pm 0.1$ instead of a scatter of 0.05 in the comparison between CHP and \archangel\ magnitude values. Attributed equally, the 0.1 scatter gives an uncertainty about $\pm 0.07$ magnitude for each source. Regardless, it is reassuring that our magnitudes are in agreements with these two alternative computations. 

As a result, these three magnitudes can be used nearly interchangeably. For a better precision they are averaged when more than one of them is available in the next sections.

\section{Robustness of the calibration of the mid-IR Tully-Fisher relation}
\label{calib}

In this section, the robustness of both the calibration method and the mid-IR TFR \citep[hereafter S13]{2013ApJ...765...94S} is shown. The 2013 calibration which was presented as preliminary, especially because of the lack of completeness of the calibrator sample, is confirmed. Magnitudes used in this section come from \archangel\ combined with a S$^4$G-pipeline or a CHP-pipeline magnitude or both when they are available. These raw magnitudes $[3.6]$ are then corrected  $[3.6]^{b,i,k,a}$ for 1) extinctions (both galactic and internal) $[3.6]^{b,i}$, 2) shift in fluxes due to Doppler effect $[3.6]^{k}$, and 3) extended emission from the Point Spread Function outer wings and from scattered diffuse emission across the IRAC focal plane $[3.6]^{a}$. These corrections are described separately in 1) \citet{1989ApJ...345..245C,1998ApJ...500..525S} and \citet{1995AJ....110.1059G,1997AJ....113...22G,1998AJ....115.2264T}, 2) \citet{1968ApJ...154...21O, 2007ApJ...664..840H} and 3) \citet{2005PASP..117..978R} and specifically for Spitzer IRAC 3.6 microns data in \citet{2012AJ....144..133S}. The resulting magnitudes are called $[3.6]^{b,i,k,a}$ in the rest of the paper where each superscript stands for a correction. 

\subsection{An updated list of galaxies}

S13 derived a template TFR using 213 galaxies in 13 clusters. The zero point calibration was given by 26 additional galaxies. The inverse fit was used to calculate the slope of the relation and a very small correction was computed to remove a bias. In this paper, the same analysis is done using an updated sample of template and zero point calibrators. This sample is improved in two aspects. The number of calibrators is increased from 213+26 to 287+32. Also galaxies are now selected in the K Band which decreases the selection bias. The selection of calibrators is extended to be complete to K=11.75 mag, the limit of the 2MRS 11.75 survey \citep{2012ApJS..199...26H}. This new set of calibrators follows the same rules as in S13: 1) candidates are chosen out of a projection-velocity window, 2) morphological types earlier than Sa are excluded, 3) HI profiles are not confused, 4) the candidates do not appear pathological, for example, exhibiting tidal disruption, and 5) inclinations must be greater than 45$^\circ$. The zero point calibrators also need to have a very well known distance from Cepheid or Tip of the Red Giant Branch measurements. There is no evidence that rejected galaxies preferentially lie in any particular part of the Tully-Fisher diagram \citep{2012ApJ...749...78T}. \\

HI linewidths are provided by the HI subproject of {\it Cosmicflows}  \citep[Extragalactic Distance Database (EDD) website\footnote{http://edd.ifa.hawaii.edu; catalog `All Digital HI'},][]{2009AJ....138.1938C,2011MNRAS.414.2005C}, Table \ref{CalibTFv2} (complete table online) gives the measurements for the calibrators.  We proceed exactly as in S13: \\

\begin{figure}
\centering
\includegraphics[scale=0.45]{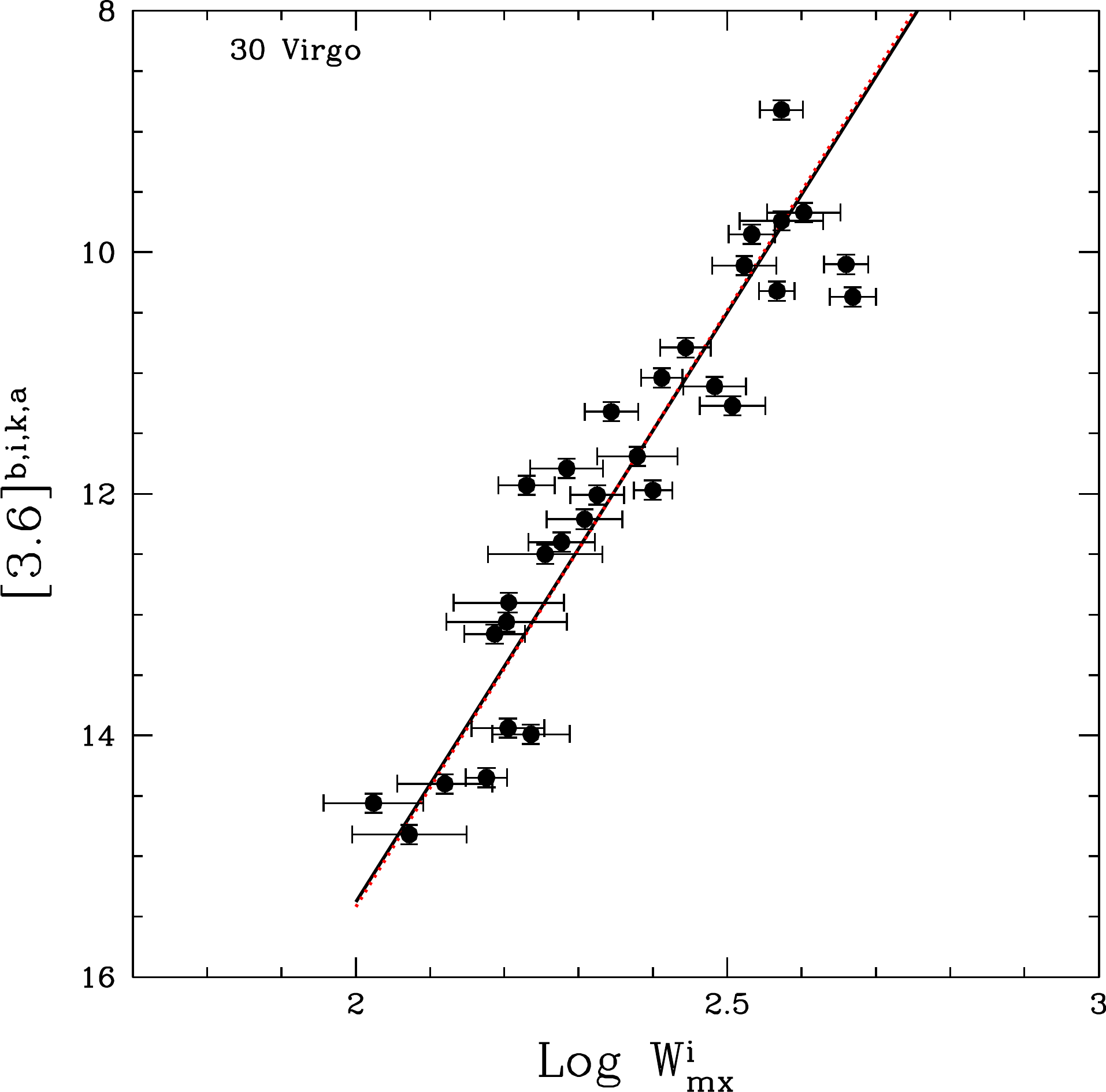}
\includegraphics[scale=0.45]{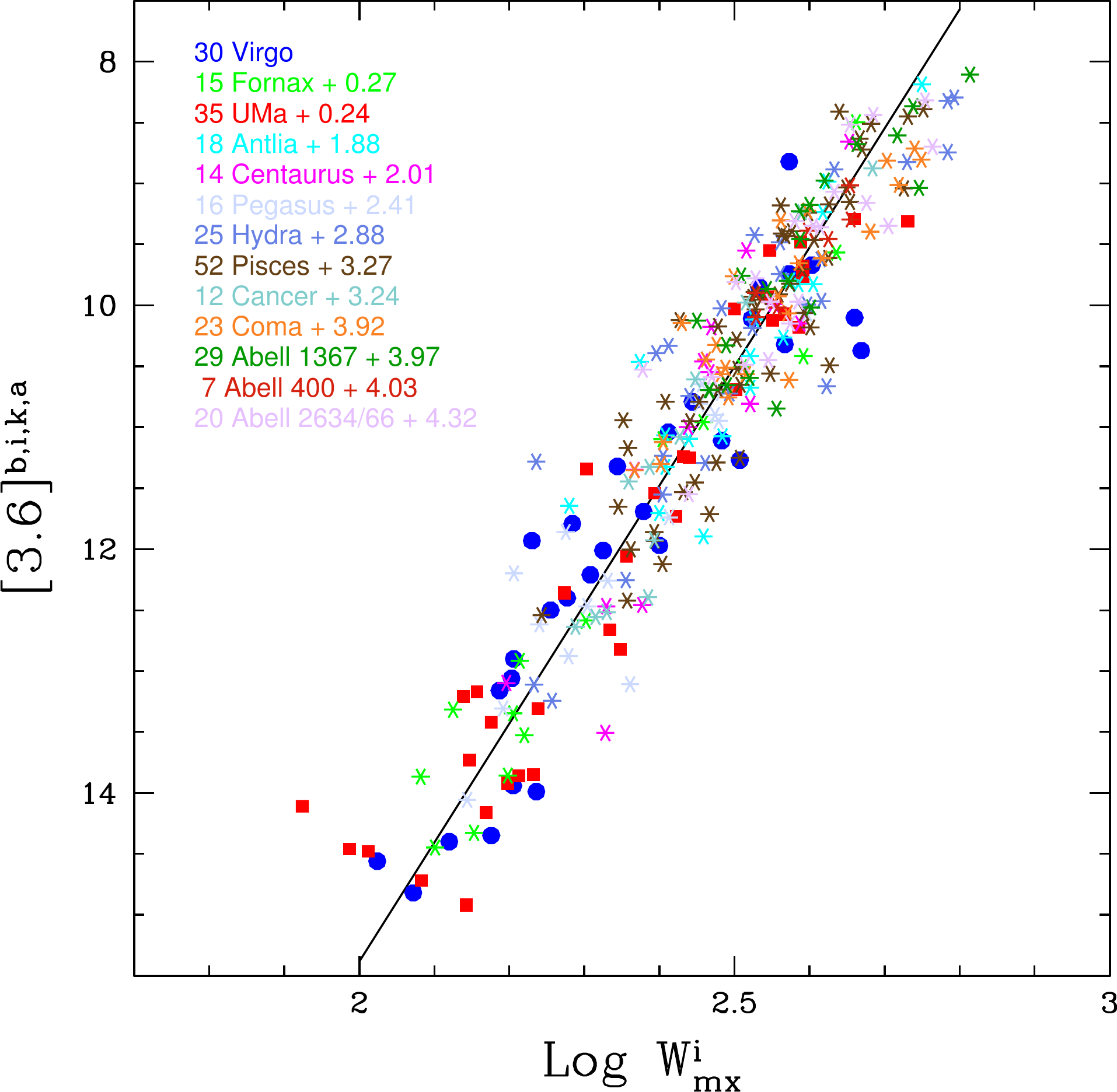}
\caption{Top: Inverse Tully-fisher relation at 3.6 microns for the Virgo cluster in dotted red line. The solid black line stands for the inverse Tully-Fisher of the template cluster. Bottom: Universal inverse TFR at 3.6 microns obtained with 287 galaxies in 13 clusters. Numbers of galaxies selected for the calibration per clusters are given in front of clusters' names while distance modulus differences between each cluster and Virgo are visible after clusters' names.}
\label{TFVir}
\end{figure}

1) An inverse TFR is fitted to each one of the clusters separately. Figure \ref{TFVir} top shows the example of the Virgo cluster. Parameters for every cluster are given in Table \ref{tbl:clfits}. The inverse fit assumes errors only in linewidth to obtain results close to free of Malmquist magnitude selection bias. Yet, there will be a tiny bias residual because of the bright end cutoff of the luminosity Schechter function although it should be somewhat smaller than with the S13 calibration where, in addition, the selection was made in the B band. We investigate this bias relic at the end of this section.\\

2) Because slopes are quite similar between clusters in Table \ref{tbl:clfits}, individual fits are consistent with the postulate of a universal TFR. Thus the 13 clusters are combined into one template cluster. Virgo is taken as the reference cluster and each one of the 12 other clusters is shifted to be on the same scale. Three by three, clusters are inserted into the template and offsets between them and Virgo are found by an iterative process which relies on least squares fits of the inverse TFR. Convergence is quick. We obtain a slope of -9.77 $\pm$ 0.19, insignificantly different from the previous slope -9.74 confirming the robustness of the S13 calibration and of the method. The universal slope and the offsets with respect to Virgo are shown on Figure \ref{TFVir} bottom. \\

\begin{figure}
\centering
\includegraphics[scale=0.44]{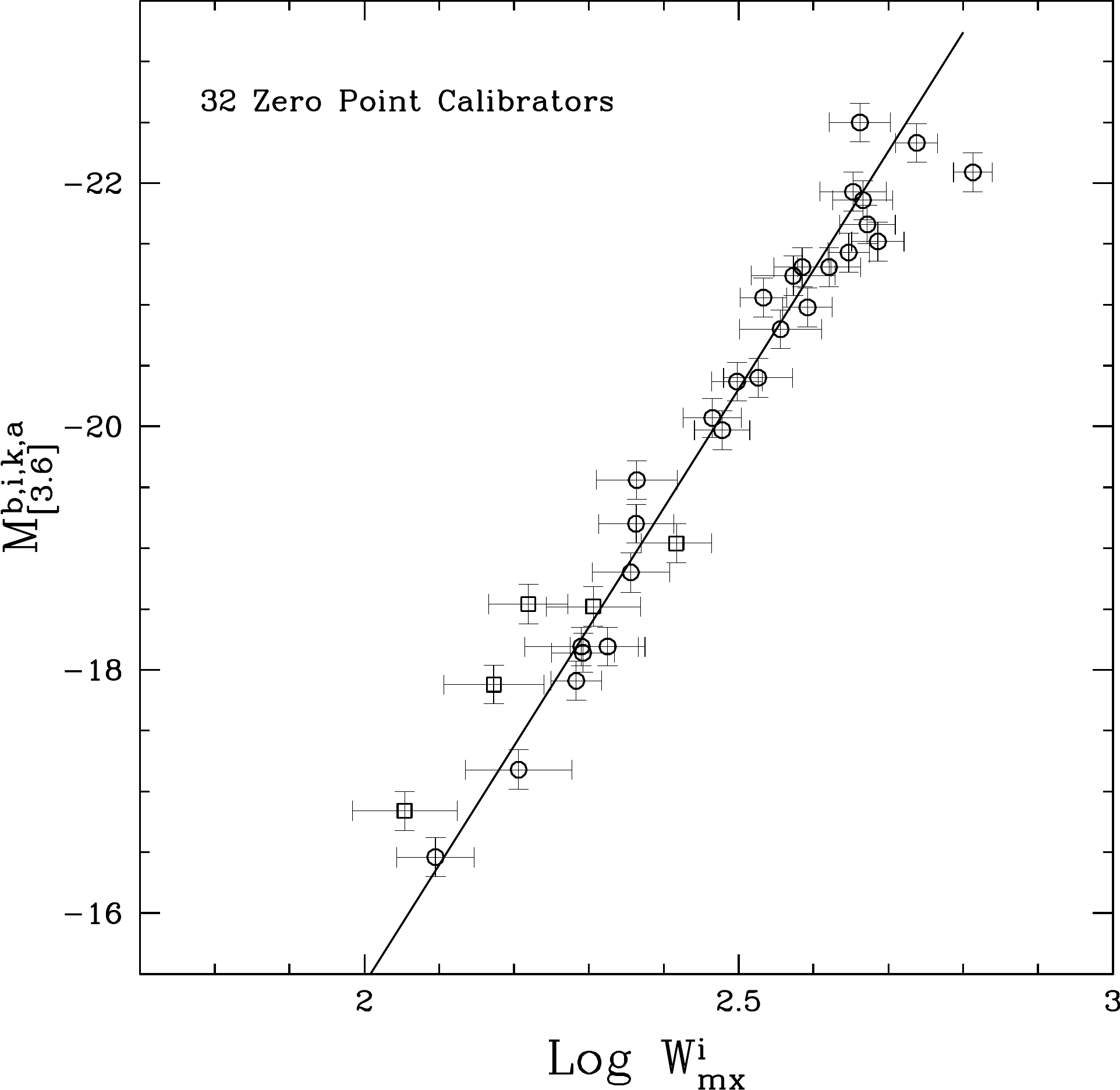}
\includegraphics[scale=0.44]{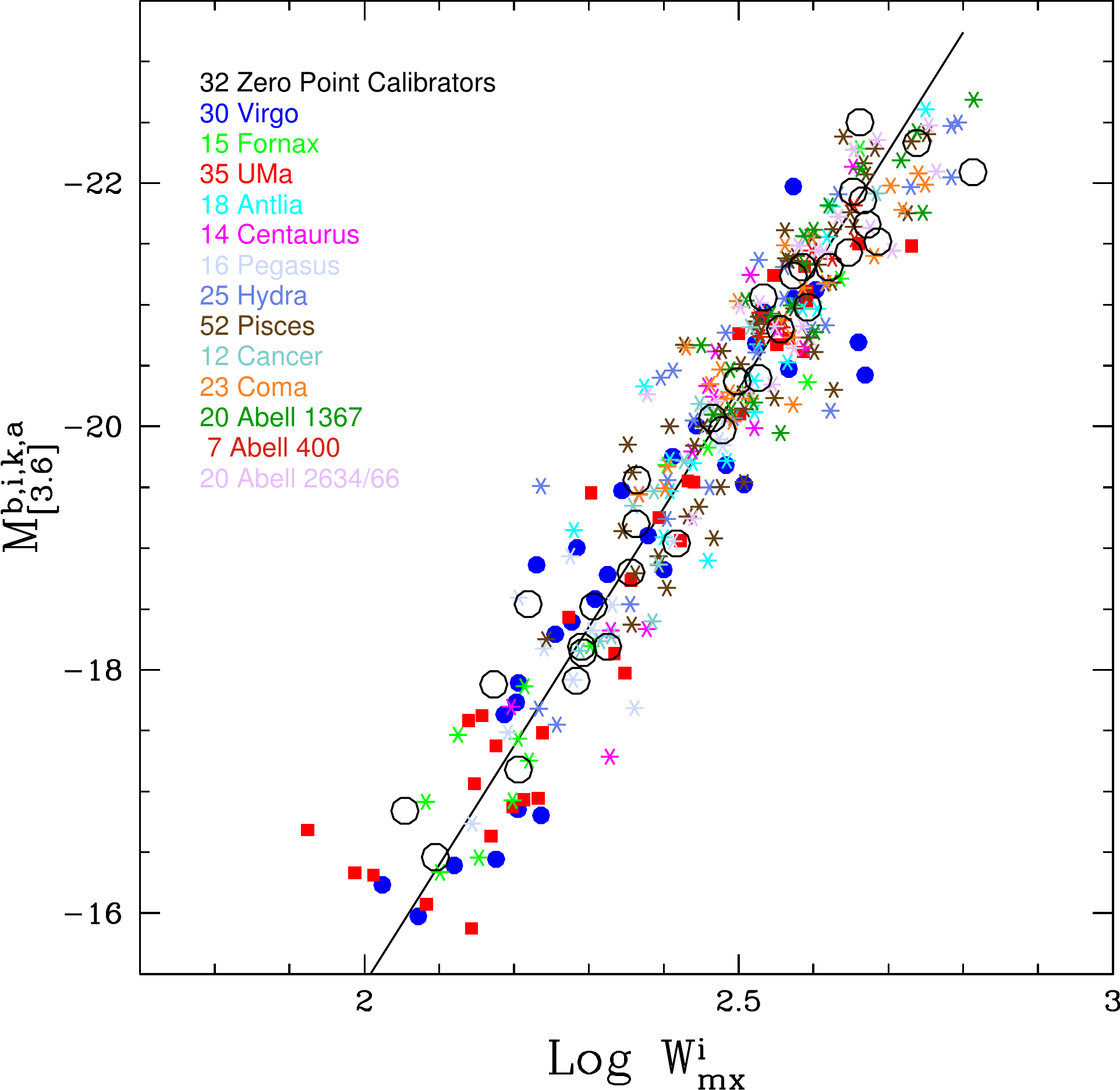}
\caption{Top: Inverse TFR for the 32 zero point calibrators with distances obtained with Cepheids (circles) or Tip of the Red Giant Branch (squares). The slope of the solid line is given by the luminosity-linewidth correlation of the template cluster while the zero point is obtained with the least squares fit to the 32 galaxies. The zero point is set at logW$^i_{mx}$ = 2.5. Bottom: Inverse Tully-Fisher relation at 3.6 microns with the slope built out of 287 galaxies in 13 clusters and the zero point set by 32 galaxies with very accurate distances.}
\label{TFZP}
\end{figure}

3) The zero point scale of the Cepheid calibrators is set by the distance modulus of the Large Magellanic Cloud, 18.48 $\pm$ [0.04-0.07] \citep{2012ApJ...759..146M,2011ApJ...730..119R}. Then, the 32 zero point calibrators give the zero point of the universal TFR assuming the slope of the cluster template. Their correlation is visible in the top panel of Figure \ref{TFZP} where now absolute magnitudes replace apparent magnitudes. The zero point of the TFR is the difference between the zero point given by zero point calibrators on Figure \ref{TFZP} top and by Virgo in Figure \ref{TFVir} top: -20.31 $\pm$ 0.09. The zero point is once again insignificantly larger than that of the S13 calibration of -20.34. \\

The universal relation at 3.6 microns is visible on Figure \ref{TFZP} and is given by a slightly updated version of the S13 calibration:
\begin{equation}
M^{b,i,k,a}_{[3.6]}=-(20.31 \pm 0.09) - (9.77 \pm 0.19)(\mathrm{log}W^i_{mx} - 2.5)
\end{equation}
with a scatter of 0.54 for the 13 clusters and 0.45 for the 32 zero point calibrators. S13 already discussed the causes of such a scatter. Among these reasons, they evoke a color term due to the fact that faster rotators tend to be redder and rise more quickly than bluer galaxies in the Tully-Fisher diagram (e.g. S13 Fig. 6). Following the earlier work, we apply a color correction in the next subsection to confirm the color corrected TF relation derived in S13.

\subsection{The color correction}

\begin{figure}
\centering
\includegraphics[scale=0.6]{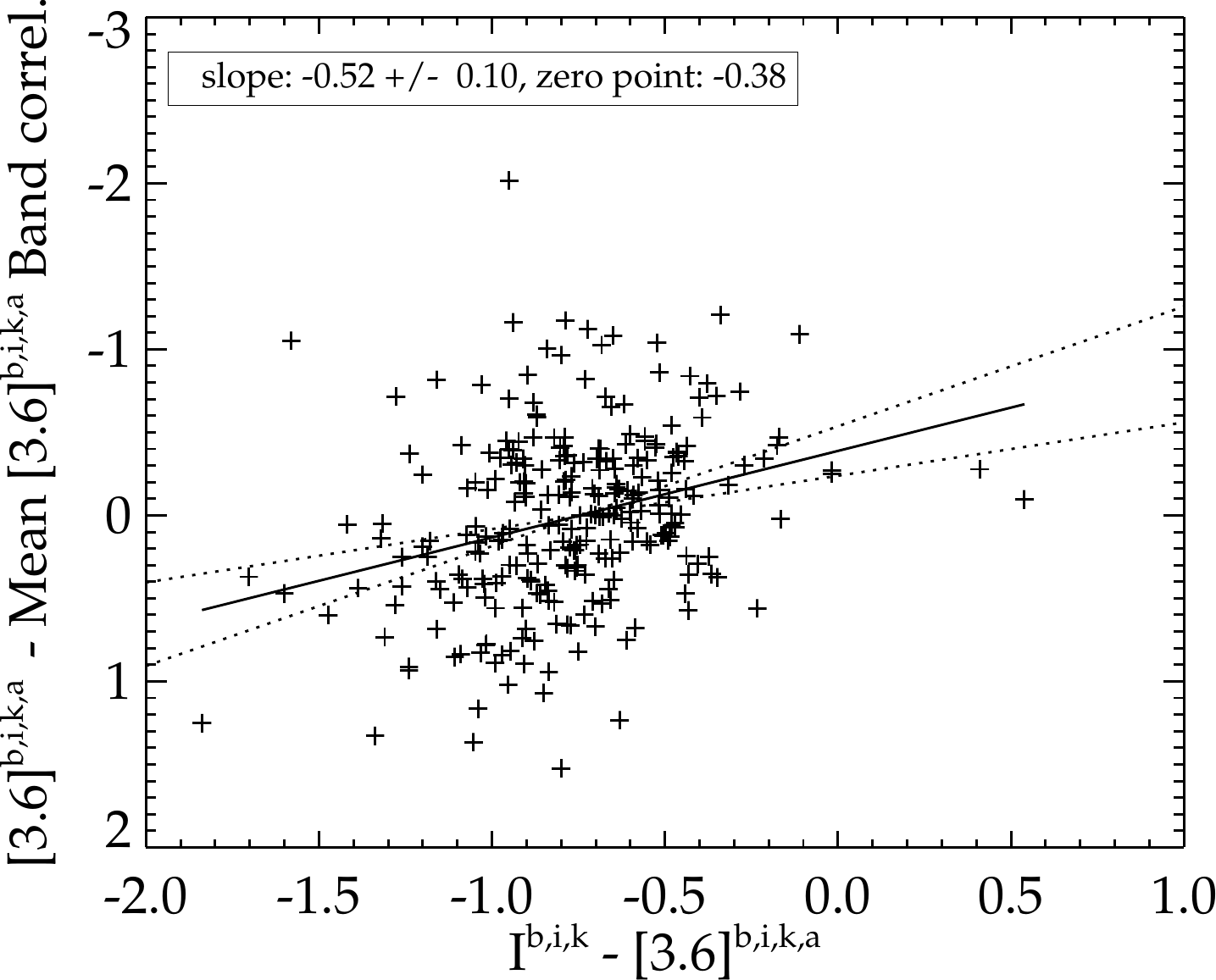}\\
\hspace{-0.2cm}\includegraphics[scale=0.45]{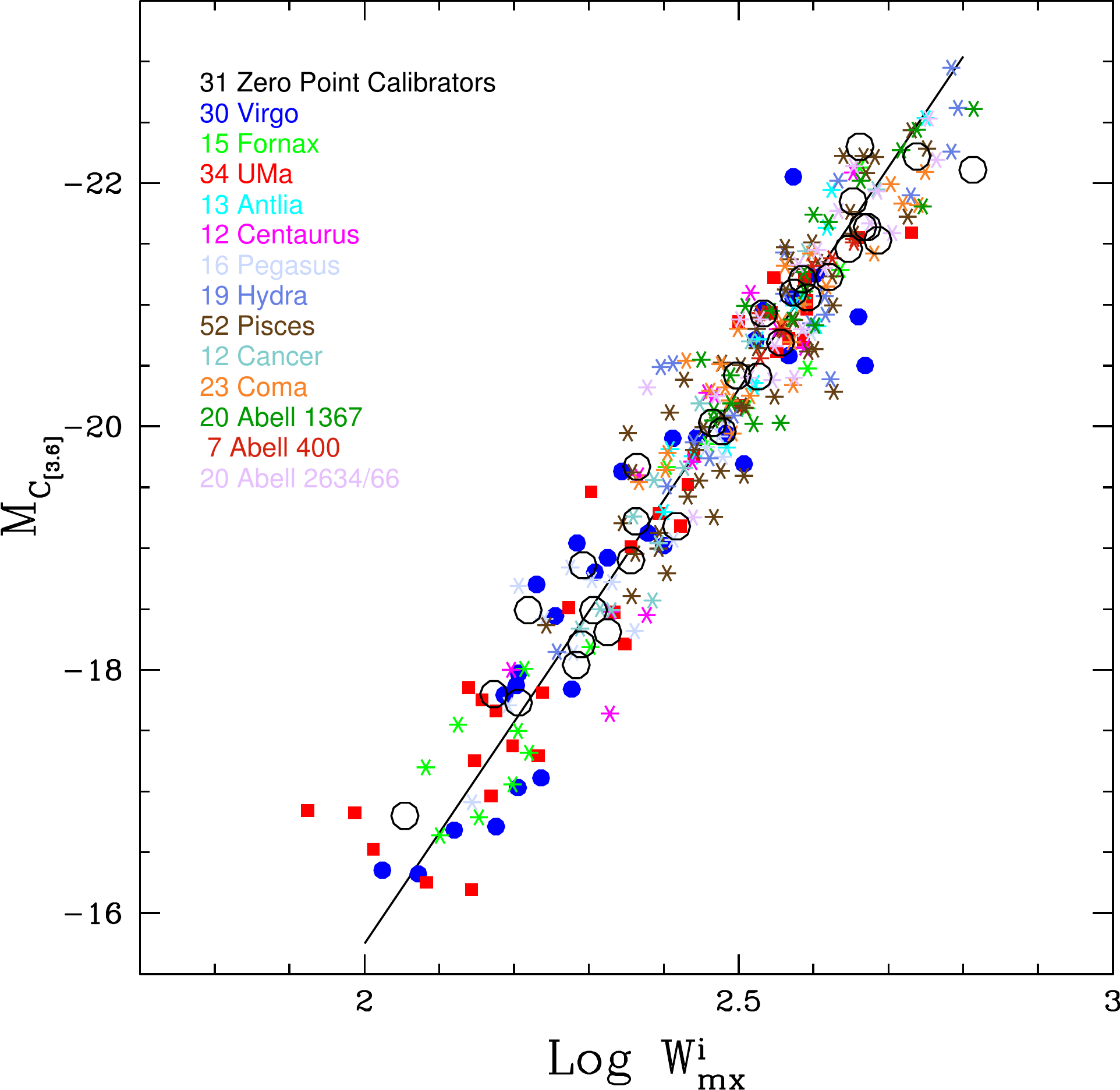}
\caption{Top: Deviation from the universal inverse TFR as a function of I$^{b,i,k}$ - [3.6]$^{b,i,k,a}$ color. The solid line stand for the best fit while the dotted lines represents the 95\% probability limits. Redder galaxies tend to lie above the relation while bluer galaxies are preferentially below the relation. Bottom: Relation for pseudo-absolute magnitudes with the zero point set by galaxies with independent very accurate distance estimates (open circles).}
\label{TFca}
\end{figure}

Because of the increased number of data, we double check the color term deriving a new estimate. The straight line fit given in Figure \ref{TFca} top is a least squares minimization with respect to the difference in magnitude of a galaxy from the derived TFR. In the [3.6] band, a galaxy is offset from the TFR by:
\begin{equation*}
\Delta M^{color}_{[3.6]}=M^{b,i,k,a} + 20.31 + 9.77(\mathrm{log}W^i_{mx} - 2.5)
\end{equation*}
\begin{equation}
\hspace{1.92cm} = -(0.52\pm0.10)[(I^{b,i,k}-[3.6]^{b,i,k,a})+0.73]
\end{equation}
Note that I Band magnitudes have been converted from the Vega to the AB system by making a 0.342 mag shift. Slope and zero point are slightly smaller than those given in S13 (-0.47 and -0.36) but within the uncertainty. Still, for completeness, we use this new estimate. Color adjusted parameters, C$_{[3.6]}$ = [3.6]$^{b,i,k,a}$-$\Delta[3.6]^{color}$, are derived accordingly and then, considered as pseudo-magnitudes to produce the color corrected calibration, proof of the robustness of the S13 calibration. The procedure described in the previous subsection is reiterated with a number of galaxies slightly decreased due to a lack of I Band measurements (273+31).\\

The color corrected calibration is visible on Figure \ref{TFca} bottom and given by:
\begin{equation}
M_{C_{[3.6]}}=-(20.31\pm0.07)-(9.10\pm0.21)(\mathrm{log}W^i_{mx}-2.5)
\end{equation}
with 0.45 and 0.37 as new scatters. A summary of the derived parameters for the TFR in this paper are given in Table \ref{tbl:clfits} as well as in Table \ref{tbl:compare} along those of S13 and those of \citet{2012ApJ...749...78T} for the I Band. Although a direct comparison has some imprecision because of the different galaxy samples, the agreement is excellent. The robustness of the procedure and of the derived TF relations is confirmed. Namely, no major bias affects the relation as it is almost independent of the calibrator sample in terms of completeness and band selection.

\subsection{Bias and distances}

Although all TFRs (individual and universal) derived in this paper are inverse fits (errors solely in linewidths), a small Malmquist selection bias residual remains. This bias was investigated with the S13 TFR calibration at 3.6 microns. In this paper, the situation is improved because galaxies are selected in K (instead of B) band. This change in wavelength selection reduces the interval between sample selection and photometry bands. However, because of the morphology of the luminosity function, galaxies are not scattered up and down exactly similarly. The amplitude of the bias increases with distance as the selection limit approaches the exponential cutoff of the luminosity function.\\

\begin{figure}
\centering
\includegraphics[scale=0.44]{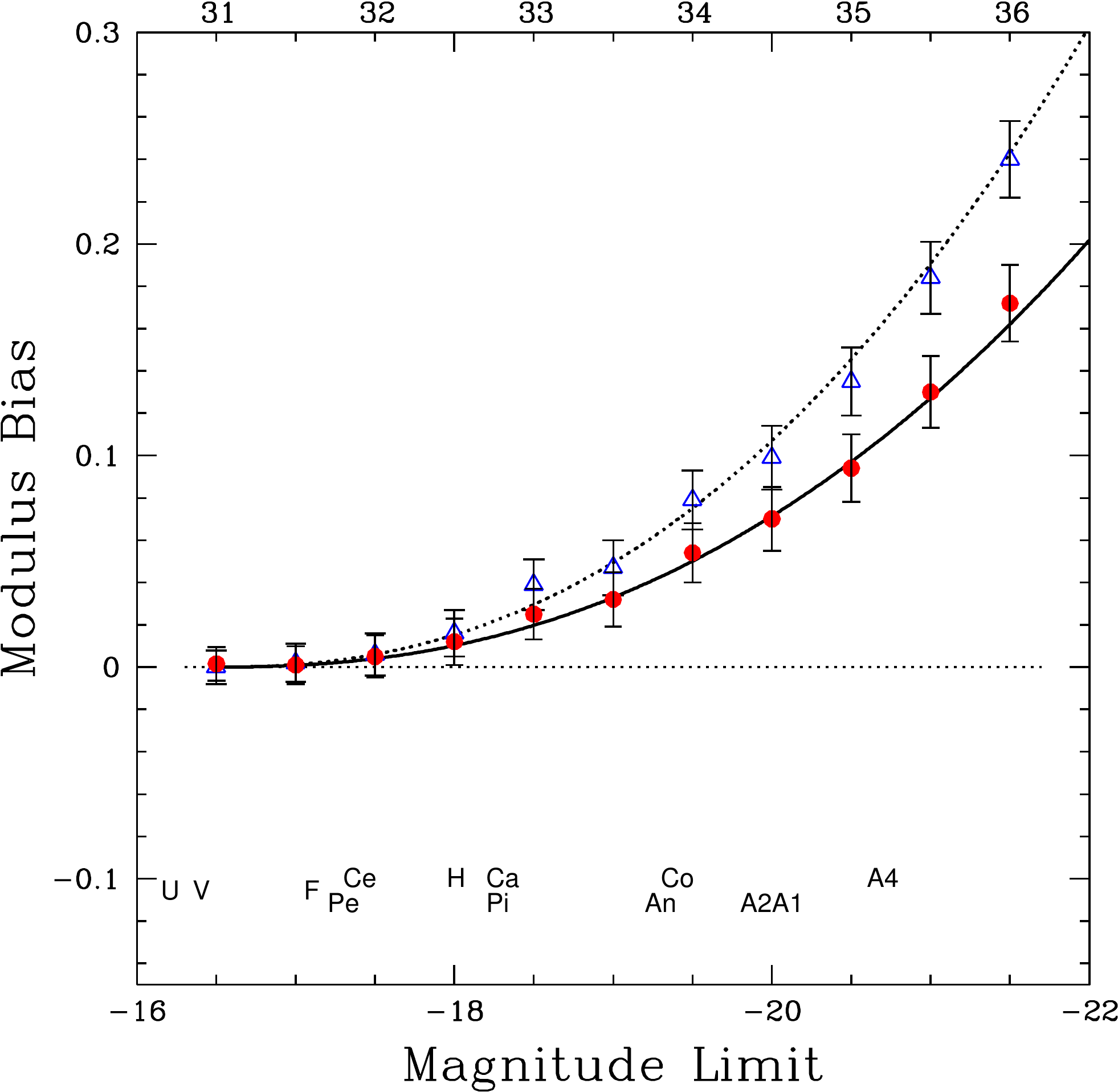}\\
\caption{Bias measured as a function of absolute magnitude cutoff. The dotted and solid black curves are fits to the blue triangles and red filled circles which are bias estimates at successive cutoffs for the [3.6] TF calibration and for the color adjusted TF relation. The formula for the curves are $0.006 (\mu-31)^{2.3}$ and $0.004 (\mu-31)^{2.3}$. Letters at the bottom stand for the 13 clusters given in Table \ref{tbl:clfits}. They are positioned at the magnitude limits of clusters and their vertical projections onto the curve give the corresponding biases. The bias for an individual galaxy with a measured modulus is given by projection onto the curves from the top axis.}
\label{bias}
\end{figure}

As a result, the same bias analysis as in S13 is conducted but without consideration of a faint end cutoff color dependence. Virgo, Fornax and Ursa Major are modeled with a \citet{1976ApJ...203..297S} function with a faint end slope of $-1.0$ and a bright end cutoff at $-22$. Then, a random population is built out of this Schechter function to match the TFR at 3.6 microns in terms of slope, zero point and scatter. The bias is estimated as the average deviation of sampled distances from the input TFR for successive brighter cutoffs with the convention, bias = input TFR - measured TFR for the different cutoff samples. The corresponding curve normalized to zero at a distance modulus of 31 is shown in Figure \ref{bias} and can be written: 
\begin{equation}
bias=0.004 (\mu-31)^{2.3}
\end{equation}
where $\mu$ is the distance modulus. The coefficient 0.004 is smaller than in S13 (0.0065) because of the previous color dependence. However, the 2.3 exponent is larger than before because of a larger assumed scatter. The scatter dominates the bias relic. At the bottom of Figure \ref{bias}, letters standing for the 13 clusters are positioned at their cutoffs while the corresponding biases are given by projection onto the curve. Bias corrections for each cluster are given in Table \ref{tbl:clfits} alongside the letters to match them with the names of clusters. Corrections are already included in moduli and distances given in this same table. As for an individual galaxy, the bias corrected distance modulus $\mu$ is obtained by adding $0.004 (\mu-31)^{2.3}$. For completeness, the bias correction for the \textit{non} color adjusted relation, obtained similarly, is given by $bias=0.006 (\mu-31)^{2.3}$. \\

\begin{figure}
\centering
\includegraphics[scale=0.46]{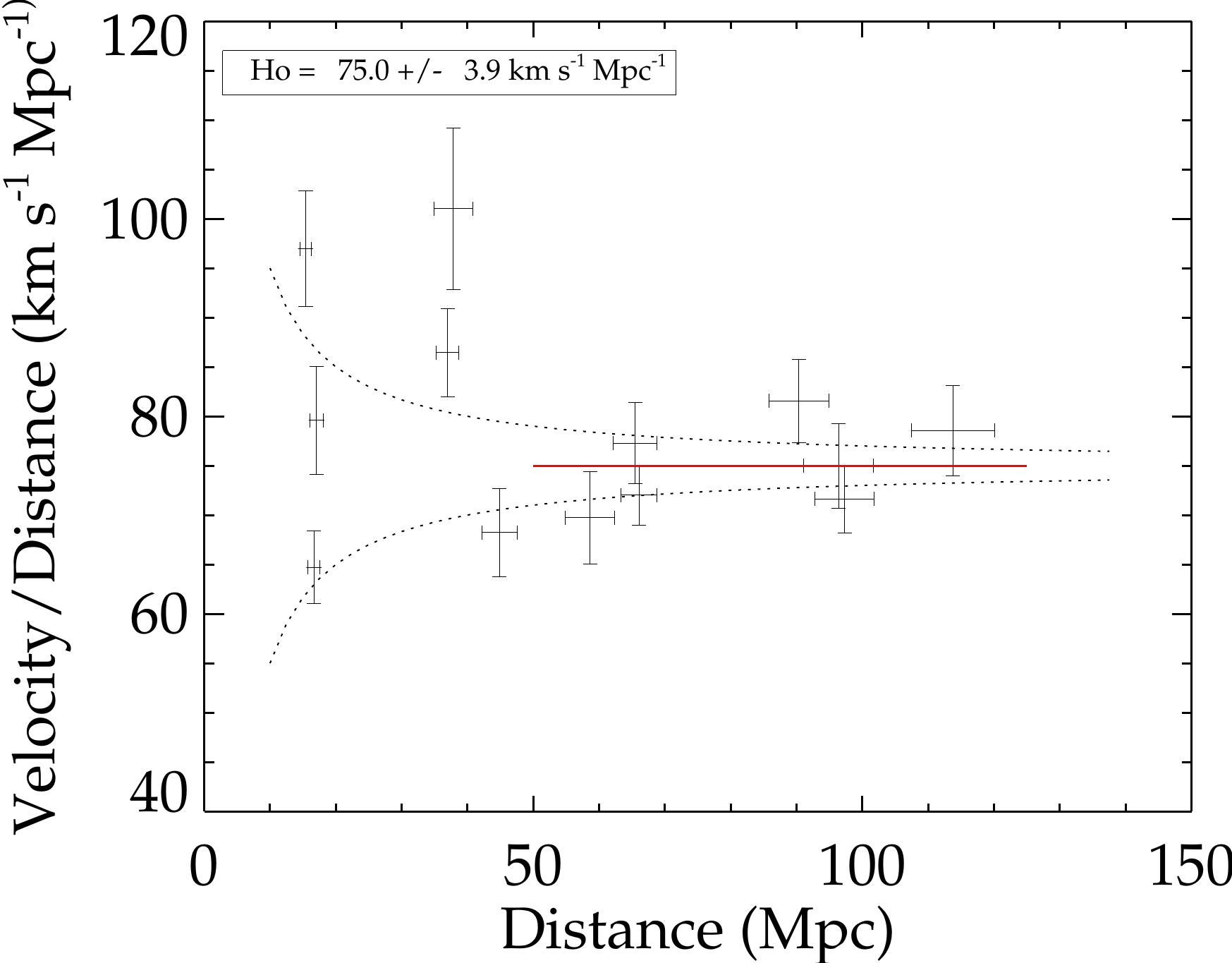}\\
\caption{Hubble parameter as a function of distance. The solid red line at 75.0 $\pm$ 3.9 km s$^{-1}$ Mpc$^{-1}$ is a fit to the logarithms of cluster "Hubble parameters" at distances greater than 50 Mpc. The dotted line represents the average 200 \kms\ deviation from the expansion due to peculiar motions.}
\label{H0}
\end{figure}

Distances obtained for the 13 clusters are compared with previous estimates (S13 and I Band) in Table \ref{tbl:compIcal}. Overall distances are in good agreement with each other and within uncertainties. Combining these distances with velocities with respect to the CMB corrected with a cosmological model assuming $\Omega_m=0.27$ and $\Omega_{\Lambda}=0.73$ \citep{2013AJ....146...86T}, it is possible to derive a "Hubble parameter" for each cluster. These values are given in Table \ref{tbl:clfits} and plotted in Figure \ref{H0}. A straight line fit to the logarithms of these parameters for clusters at a distance greater than 50 Mpc gives a Hubble value of 75 $\pm$ 4 (ran) km s$^{-1}$ Mpc$^{-1}$ where (ran) stands for twice the 1-$\sigma$ random error.

\section{Confirming Hubble Constant estimate with Supernovae}
\label{snIa}

\begin{figure}
\centering
\includegraphics[scale=0.43]{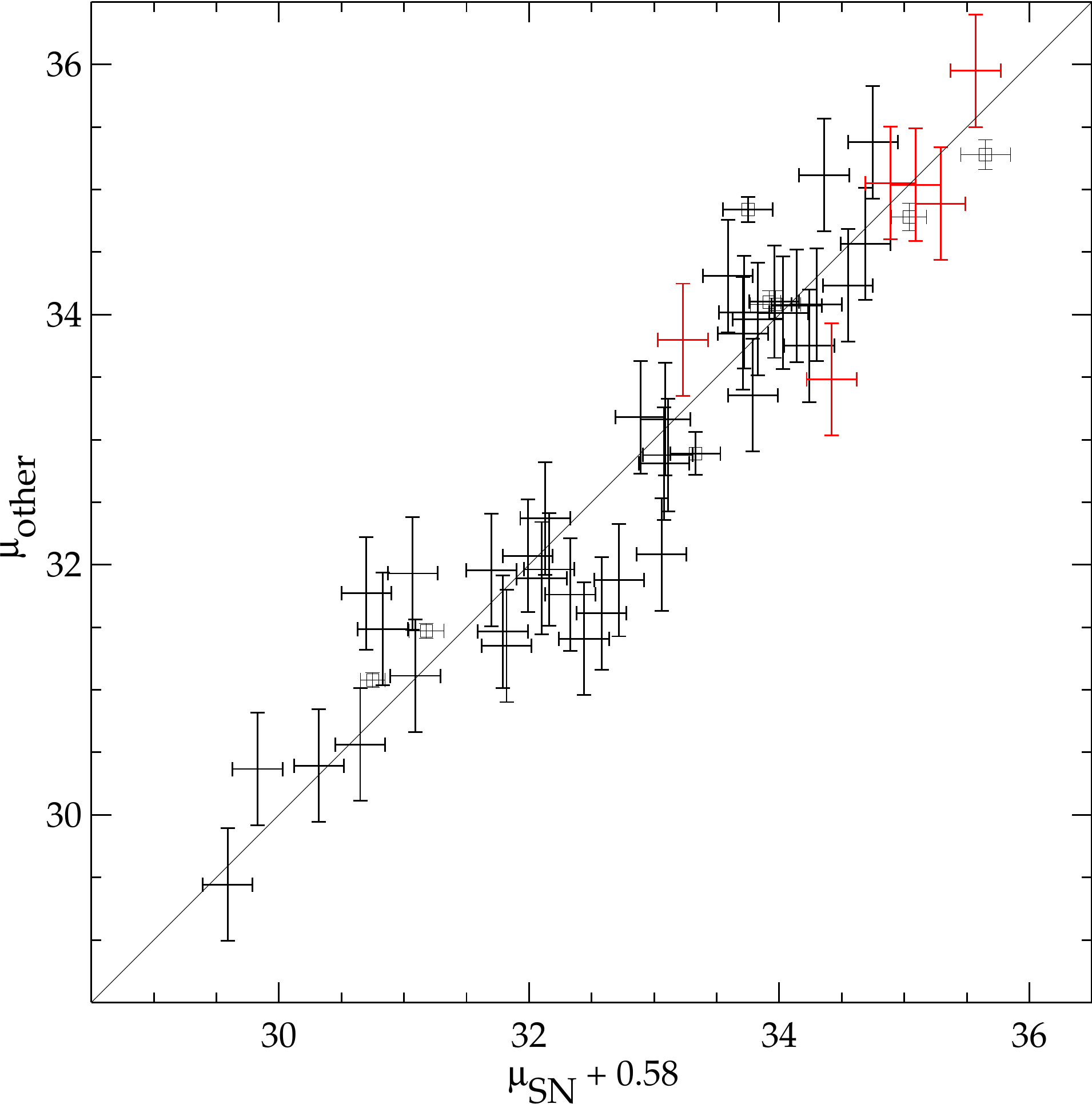}\\
\caption{Top: Comparison between moduli derived with SNIa and with "other" methods (TFR, with Cepheid and Surface Brightness Fluctuation supplements).  The solid line has for slope the weighted fit to the 45 galaxies (filled points) with TFR distances and six of the eight clusters (open squares) and a null zeropoint. }
\label{snia}
\end{figure}

At the time of \citet{2012ApJ...758L..12S} only 39 hosts of SNIa had been observed with Spitzer. Now, 45 host galaxies have all the required parameters to be compared with SNIa measurements. The new information extends the previous work by only six galaxies and we do not expect much change with regard to the offsets between SNIa and TF distance moduli estimates, nor [3.6] band measurements especially because the calibration at 3.6 $\mu$m has been shown to be very robust. Still, for the sake of completeness, raw magnitudes of these galaxies are corrected as before and the corresponding pseudo-magnitudes are derived. The color corrected TFR is applied to this set of supernova hosts to derive distance moduli estimates. These distance moduli are then bias corrected and compared with distance moduli obtained from supernova measurements to determine the supernova zero point scale.  All the parameters are gathered in Table \ref{tblCh3:SNIav2}. Figure \ref{snia} shows the results when the six additional galaxies are included in the sample and the TFR is used to derive moduli. Eight of the thirteen calibration clusters with observed SNIa are also added. The straight line 
is a fit, assuming slope unity, to the 45 individual galaxies each with weight 1 and six clusters each with weight 9 (Centaurus and Abell 1367 have been rejected in \citet{2012ApJ...758L..12S} and distance moduli for Virgo and Fornax include contributions from Cepheid and Surface Brightness Fluctuation methods for consistency with the previous work). The offset is identical to that found in S13. Our Hubble Constant estimate is unchanged  H$_0 = 75.2 \pm 3.3$~\kmsMpc.

\section{A catalog of accurate distance estimates}
\label{vrad}

This paper has been the occasion to release the observational campaign Cosmicflows with Spitzer (CFS), a photometric component of the Cosmicflows project. The primarily goal of this observational survey is to increase the number of distance estimates close to the Zone Of Avoidance using the Tully-Fisher relation. The first channel (3.6 $\mu$m) of the InfraRed Array Camera onboard the Spitzer Space Telescope is indeed the instrument of choice to obtain the required excellent photometry.  At this wavelength the Zone of Avoidance and uncertainties on measurements are considerably reduced. Surface photometry of 1270 galaxies constituting the CFS sample observed in cycle 8 with IRAC channel 1 and over 400 additional galaxies observed in various other surveys have been presented in sections \ref{observation} and \ref{reduction}. The Spitzer Survey of Stellar Structure in Galaxies supplies many more galaxies of interests to the Cosmicflows project.

 The final set is constituted of 1935 galaxies with required parameters (in particular W$_{mx}$, b/a, [3.6] and if available I magnitudes), to derive an estimate of their distance with the mid-infrared (color adjusted) TFR derived in section \ref{calib}, all available. Axial ratios come either from previous estimates of the Cosmicflows program or from HyperLeda if they are from \citet{2003A&A...412...45P}. I-band magnitudes come from a multitude of surveys set on the same scale. The compilation of I band magnitudes is described in \citet{2013AJ....146...86T}. It gathers magnitudes used in \citet{2000ApJ...533..744T,2008ApJ...676..184T}, themselves borrowing from \citet{1997AJ....113...22G,1992ApJS...81..413M,1988ApJ...330..579P,1996AJ....112.2471T}, but also recent derivations from \citet{2011MNRAS.415.1935C,2007ApJS..172..599S} and \citet{2012MNRAS.425.2741H}. \citet{2013AJ....146...86T} showed that these I-band magnitudes are on a consistent scale after small adjustments with the exception of those of \citet{2012MNRAS.425.2741H} because they use a significantly different filter. Accordingly these later are corrected with the formulas prescribed by \citet{2002AJ....123.2121S} and \citet{2013AJ....146...86T}. These corrections involve a translation from Sloan g, r, i band (Gunn i band) to Cousins I band:
 \begin{equation}
 I_{sdss}^c = i-0.14(g-r)-0.35 
  \end{equation} 
  where cases with r-i$\ge$0.95 are excluded, and account for a slight tilt between $I_{sdss}^c$ and $I_c$, from the Cosmicflows project, magnitudes.
  \begin{equation}
 I_c = 1.017 \;I_{sdss}^c-0.221
 \end{equation}
 I-band magnitudes are extinction and k-corrected with the formulas given in \citet{2010MNRAS.405.1409C,2000ApJ...533..744T}. Then I-band magnitudes are converted to the AB system. [3.6] magnitudes are also corrected and pseudo-magnitudes are derived. Combined with the (color corrected) Tully-Fisher relation applied to linewidths, these latter enable the derivation of distance moduli. Distance moduli are corrected for the selection bias before deriving distance estimates. Table 6 gives the first few derived distance estimates. Eventually these distance estimates will be incorporated into a new data release of the Cosmicflows project, increasing the size of the previous catalog by 20\%, including spatial regions close to the Zone Of Avoidance.


\section{Conclusion}

With the new generation of sensitive telescopes/detectors both in the radio band and in the photometric domain, cosmic flow studies have received an impetus. The space-base Spitzer telescope is an example of such a telescope with enhanced capacities. With a Spitzer-adapted version of the software \archangel, we have obtained surface brightness photometry and distances for 1270 galaxies that are part of the {\it Cosmicflows with Spitzer} program, itself included in the larger {\it Cosmicflows} project. An increase in the number of Tully-Fisher calibrators since the 2013 calibration and a superior selection criteria using K-band instead of B-band led us to recalibrate the 3.6 micron TF relation. The derived relation confirms the robustness of the 2013 calibration and is given by $M_{C_{[3.6]}}=-(20.31\pm0.07)-(9.10\pm0.21)(\mathrm{log}W^i_{mx}-2.5)$ with a scatter of 0.43 mag ($\sim$ 22\% in distance). $M_{C_{[3.6]}}$ is the pseudo magnitude obtained after correction of [3.6] magnitudes by I-[3.6] colors, $M_{C_{[3.6]}} = M_{[3.6]}^{b,i,k,a}+(0.52\pm0.10)[(I^{b,i,k}-[3.6]^{b,i,k,a})+0.73]$ where I Band magnitude have been shifted to the AB system. Resulting distance moduli $\mu$ are then corrected for a tiny bias effect with the addition of the term $0.004 (\mu-31)^{2.3}$. Applying this calibration to a set of supernova hosts to obtain a scale for the supernovae, we confirm our Hubble Constant estimate 75.2 $\pm$ 3.3 km s$^{-1}$ Mpc$^{-1}$. \\

Drawing from the Spitzer archive, consistent magnitudes are available for 1935 galaxies that also have suitable HI linewidth measurements and appropriate morphologies and inclinations for the determination of TFR distances.  This new material substantially augments the compilation of distances and derivative peculiar velocities in the {\it Cosmicflows} program.  The all-sky uniformity of the satellite photometry mitigates concerns that spatially correlated errors might induce artificial flows and the observations in the mid-infrared negate concerns with reddening even at low galactic latitudes.  A parallel program using mid-infrared data from the Wide-Field Infrared Survey Explorer (WISE) complements the present study \citep{2014Neill}.  Together, the new distances will make a major contribution to what will become Cosmicflows-3 and further enable reconstructions of local structure \citep{2014Tully} and constrained simulations of the development of that structure \citep{2014MNRAS.437.3586S}.

\section*{Acknowledgements}
The data used in this paper are available at the Extragalactic Distance Database. We especially thanks our {\it Cosmicflows with Spitzer} collaborators Wendy Freedman, 
Barry Madore, 
Eric Persson and Mark Seibert. 
We are indebted to James Schombert for the development of the \archangel\ software and we thank him for his useful comments as a referee. We thank Kartik Sheth for discussions regarding Spitzer photometry. NASA through the Spitzer Science Center provides support for CFS, {\it Cosmicflows with Spitzer}, cycle 8 program 80072. This research has made use of the NASA/IPAC Extragalactic Database (NED) which is operated by the Jet Propulsion Laboratory, California Institute of Technology, under contract with the National Aeronautics and Space Administration. We acknowledge the usage of the HyperLeda database (http://leda.univ-lyon1.fr). HC and JS acknowledge support from the Lyon Institute of Origins under grant ANR-10-LABX-66 and from CNRS under PICS-06233. RBT acknowledges support from the US National Science Foundation award AST09-08846. TJ acknowledges the Astronomy Department of the University of Cape Town.

\begin{table*}
\begin{center}
\begin{tabular}{|l|c|c|c|r|c|c|c|c|c|c|r|r|}
\hline
Cluster & $V_{mod}$ & eV & N &Slope& ZP & rms & ZP$_{color}$ & rms & bias & DM & Dist & $V/D$ \\
\hline
V Virgo & 1495 & 37 & 30-30 & -9.88 $\pm$ 0.73 &10.50 $\pm$  0.12 & 0.64 &  10.63 $\pm$ 0.10 & 0.55 & 0.00 & 30.94 $\pm$ 0.12 &15.4 $\pm$ 0.9 & 97.0 $\pm$ 5.9\\ 
F Fornax & 1358 & 45 &  15-15 & -9.56 $\pm$ 0.63 & 10.77 $\pm$ 0.12 & 0.46 & 10.85 $\pm$ 0.11 & 0.42 & 0.00 & 31.16 $\pm$ 0.13 &17.1 $\pm$ 1.0 & 79.6 $\pm$ 5.4\\
U U Ma & 1079 & 14 & 35-34 & -9.32 $\pm$ 0.52 & 10.74 $\pm$ 0.11 & 0.64 & 10.80 $\pm$ 0.10 & 0.57 & 0.00 & 31.11 $\pm$ 0.12 & 16.7 $\pm$ 0.9 & 64.7 $\pm$ 3.7\\
An Antlia & 3198 & 74 & 18-13 & -10.07 $\pm$ 1.33 & 12.37 $\pm$ 0.12 & 0.52 & 12.48 $\pm$ 0.07 & 0.27 & 0.05 & 32.84 $\pm$ 0.10 & 37.0 $\pm$ 1.7 & 86.5 $\pm$ 4.5\\
Ce Cen30 & 3823 & 82 & 14-12 & -12.92 $\pm$ 1.74 & 12.51 $\pm$ 0.16 & 0.60 & 12.58 $\pm$ 0.16 & 0.55 & 0.00 & 32.89 $\pm$ 0.17 & 37.8 $\pm$ 3.0 & 101.0 $\pm$ 8.2\\
Pe Pegasus & 3062 & 78 & 16-16 & -9.84 $\pm$ 1.03 & 12.91 $\pm$ 0.14 & 0.55 & 12.94 $\pm$ 0.11 & 0.44 & 0.01 & 33.26 $\pm$ 0.13 & 44.9 $\pm$ 2.7 & 68.2 $\pm$ 4.4\\
H Hydra & 4088 & 72 & 25-19 & -9.12 $\pm$ 0.94 & 13.38 $\pm$ 0.14 & 0.71 & 13.52 $\pm$ 0.13 & 0.55 & 0.01 & 33.84 $\pm$ 0.14 & 58.6 $\pm$ 3.8 & 69.7 $\pm$ 4.7\\
Pi Pisces & 4759 & 39 & 52-52 & -11.02 $\pm$ 0.75 & 13.77 $\pm$ 0.07 & 0.50 & 13.76 $\pm$ 0.06 & 0.45 & 0.03  & 34.10 $\pm$ 0.09 & 66.1 $\pm$ 2.7 & 72.0 $\pm$ 3.0\\
Ca Cancer & 5059 & 82 & 12-12 & -11.65 $\pm$ 1.02 & 13.74 $\pm$ 0.11 & 0.39 & 13.75 $\pm$ 0.10 & 0.31 & 0.02& 34.08 $\pm$ 0.11 & 65.5 $\pm$ 3.3 & 77.3 $\pm$ 4.1\\
Co Coma & 7370 & 76 & 23-23 & -7.97 $\pm$ 0.67 & 14.42 $\pm$ 0.10 & 0.49 & 14.40 $\pm$ 0.09 & 0.42 & 0.07 & 34.78 $\pm$ 0.11 & 90.4 $\pm$ 4.6 & 81.6 $\pm$ 4.2\\
A4 A400 & 7228 & 97 & 7-7 & -8.00 $\pm$ 1.38 & 14.47 $\pm$ 0.11 & 0.48 & 14.46 $\pm$ 0.09 & 0.42 & 0.15 & 34.92 $\pm$ 0.12 & 96.4 $\pm$ 5.3 & 75.0 $\pm$ 4.3\\ 
A1 A1367 & 6969 & 93 & 20-20 & -9.32 $\pm$ 0.92 & 14.53 $\pm$ 0.08 & 0.21 & 14.53 $\pm$ 0.07 & 0.19 & 0.10 & 34.94 $\pm$ 0.10 & 97.3 $\pm$ 4.5 & 71.6 $\pm$ 3.4\\
A2 A2634/66 & 8938 & 164 & 20-20 & -9.55 $\pm$ 0.97 & 14.82 $\pm$ 0.11 & 0.50 & 14.88 $\pm$ 0.10 & 0.43 & 0.09 & 35.28 $\pm$ 0.12 & 113.8 $\pm$ 6.3 & 78.6 $\pm$ 4.6\\
\hline
\end{tabular}
\end{center}
\caption{Properties of the Cluster Fits: (1) Cluster name, (2) Mean velocity of the cluster with respect to the CMB corrected for cosmological effects, \kms , (3) Error on the velocity, \kms, (4) Number of studied galaxy per cluster for the original TFR and for the color-corrected TFR, (5) Slope of the inverse fit, (6) Zero point relative to Virgo's zero point, no color adjustment, mag, (7) Scatter, no color adjustment, (8) Zero point relative to Virgo's zero point after color adjustment, mag, (9) Scatter after color adjustment, mag, (10) Bias, mag, (11) Bias corrected Distance Modulus, mag, (12) Cluster Distance, Mpc, (13) Hubble parameter, km~s$^{-1}$~Mpc$^{-1}$}
\label{tbl:clfits}
\end{table*}

\begin{table*}
\begin{center}
\begin{tabular}{|l|c|c|c|c|}
\hline
 Sample                 & Ngal  & Slope  & RMS & Zero Point \\
\hline
$I$ template         &   267 & -8.81$\pm$0.16 & 0.41 &  -- \\
$I$ zero point       &    36  &       --      &   0.36 &  -21.39$\pm$0.07 (Veg) \\
2013 $[3.6]$ template   &  213 & -9.74$\pm$0.22 & 0.49 &  -- \\
2013 $[3.6]$ zero point &    26 &        --     &    0.44 &  -20.34$\pm$0.10 (AB) \\
2013 $M_C$ template  &  213 & -9.13$\pm$0.22 & 0.44 &  -- \\
2013 $M_C$ zero point &   26  &       --     &    0.37 &  -20.34$\pm$0.08 (AB) \\
This paper $[3.6]$ template   &  287 & -9.77$\pm$0.19 & 0.54 &  -- \\
This paper $[3.6]$ zero point &    32 &        --     &    0.45 &  -20.31$\pm$0.09 (AB) \\
This paper $M_C$ template  &  273 & -9.10$\pm$0.21 & 0.45 &  -- \\
This paper $M_C$ zero point &   31  &       --     &    0.37 &  -20.31$\pm$0.07 (AB) \\
\hline
\end{tabular}
\caption{TFR parameters in \citet{2012ApJ...749..174C} for the I Band obtained with the B band selected calibrator sample, in S13 for the 2013 [3.6] calibration derived with part of the B band selected calibrator sample and in this paper for the calibration computed with the K band selected calibrator sample.}
\label{tbl:compare}
\end{center}
\end{table*}

\begin{table*}
\begin{center}
\begin{tabular}{|c|c|c|c|c|c|c|c|}
\hline
Cluster & This Paper & 2013 Paper & TC12 & Cluster & This Paper & 2013 Paper & TC12\\
\hline
V~Virgo & 15.4 $\pm$ 0.9 &14.7 $\pm$ 0.9 & 15.9 $\pm$ 0.8 & Pi~Pisces & 66 $\pm$ 3 & 65 $\pm$ 3 & 64 $\pm$ 2 \\
F~Fornax &17.1 $\pm$ 1.0 & 17.4 $\pm$ 1.2 & 17.3 $\pm$ 1.0 & Ca~Cancer&66 $\pm$ 3 & 67 $\pm$ 4 & 65 $\pm$ 3 \\
U~U Ma &16.7 $\pm$ 0.9 &18.0 $\pm$ 0.9 & 17.4 $\pm$ 0.9 & Co~Coma&  90 $\pm$ 5& 95 $\pm$ 6 & 90 $\pm$ 4\\ 
An~Antlia &37 $\pm$ 2 & 37 $\pm$ 2 & 37 $\pm$ 2 & A4~A400 &96 $\pm$ 5& 97 $\pm$ 5 & 94 $\pm$ 5\\  
Ce~Cen30 &38 $\pm$ 3& 39 $\pm$ 4 & 38 $\pm$ 3 & A1~A1367 &97 $\pm$5 &  96 $\pm$ 6 & 94 $\pm$ 5\\
Pe~Pegasus &45 $\pm$ 3& 45 $\pm$ 3 & 43 $\pm$ 3 & A2~A2634/66& 114 $\pm$ 6& 112 $\pm$ 7 & / \\
H~Hydra & 59 $\pm$ 4&56 $\pm$ 4 & 59 $\pm$ 4 & A2634 && / & 121 $\pm$ 7 \\
\hline
\end{tabular}
\end{center}
\caption{Comparison with S13 and \citet{2012ApJ...749..174C}: (1) Cluster name, (2) this paper distance, Mpc, (3) S13 distance, Mpc  (4)  \citet{2012ApJ...749..174C} distance, Mpc}
\label{tbl:compIcal}
\end{table*}

\begin{table*}
\begin{center}
\begin{tabular}{|l|r|c|c|c|r|r|c|c|c|}
\hline
Name& PGC& $v_{CMB}$ & $v_{mod}$ & $W_{mx}^i$ & $[3.6]^{b,i,k,a}$& $C_{[3.6]}$ & $M_{C_{[3.6]}}$ & $\mu_{TF}$ & $\mu_{SN}$ \\
\hline
 \textcolor{red}{UGC00139}  &     \textcolor{red}{963} & \textcolor{red}{3975} & \textcolor{red}{3626} & \textcolor{red}{311} & \textcolor{red}{13.43} & \textcolor{red}{13.51} & \textcolor{red}{-20.25} & \textcolor{red}{33.80} & \textcolor{red}{33.33}\\
UGC00646                              &   3773 &5348 &4898 &389 &12.85 &12.84 &-21.13 &34.02 &33.82\\
PGC005341                            &   5341 &1964 &1601 &236 &12.84 &12.72 &-19.15 &31.88 &32.82\\
NGC0673                                &   6624 &5241 &5051 &444 &11.96 &12.15 &-21.66 &33.85 &33.81\\
NGC0958                                &   9560 &5732 &5623 &592 &11.09 &11.23 &-22.79 &34.08 &34.40\\
 \textcolor{red}{UGC01993}  &    \textcolor{red}{9618} & \textcolor{red}{8005} & \textcolor{red}{7967} & \textcolor{red}{485} & \textcolor{red}{12.97} & \textcolor{red}{12.94} & \textcolor{red}{-22.00} & \textcolor{red}{35.04} & \textcolor{red}{35.19}\\
 \textcolor{red}{IC1844}          &   \textcolor{red}{10448} & \textcolor{red}{6846} & \textcolor{red}{6693} & \textcolor{red}{309} & \textcolor{red}{13.55} & \textcolor{red}{13.23} & \textcolor{red}{-20.22} & \textcolor{red}{33.48} & \textcolor{red}{34.52}\\
ESO300-009                           &  11606 &6045 &6017 &321 &14.67 &14.64 &-20.37 &35.12 &34.46\\
 \textcolor{red}{PGC011767} &   \textcolor{red}{11767} & \textcolor{red}{8701} & \textcolor{red}{8671} & \textcolor{red}{422} & \textcolor{red}{13.16} & \textcolor{red}{13.35} & \textcolor{red}{-21.46} & \textcolor{red}{34.89} & \textcolor{red}{35.39}\\
  NGC1448       &  13727 &1194 &1062 &388 & 9.97 & 9.99 &-21.12 &31.11 &31.19\\
 UGC03329                             &  17509 &5253 &5668 &524 &11.74 &11.65 &-22.31 &34.01 &34.13\\
 UGC03375                             &  18089 &5783 &5879 &534 &11.63 &11.67 &-22.38 &34.10 &34.06\\
 PGC018373                           &  18373 &2168 &2281 &239 &12.45 &12.54 &-19.22 &31.76 &32.43\\
 UGC03432                             &  18747 &4996 &5080 &289 &13.93 &13.96 &-19.96 &33.96 &33.93\\
 UGC03576                             &  19788 &5966 &6009 &392 &12.94 &13.01 &-21.17 &34.23 &34.65\\
 UGC03770                             &  20513 &6378 &6646 &371 &13.48 &13.55 &-20.95 &34.57 &34.79\\
 UGC03845                             &  21020 &3034 &3166 &257 &13.33 &13.36 &-19.50 &32.88 &33.21\\
 NGC2841                               &  26512 & 637 & 810 &650 & 8.63 & 8.61 &-23.16 &31.77 &30.80\\
 NGC3021                               &  28357 &1515 &1781 &302 &11.64 &11.82 &-20.14 &31.96 &32.26\\
 NGC3294                               &  31428 &1567 &1838 &431 &10.76 &10.82 &-21.54 &32.37 &32.23\\
 NGC3368                               &  32192 & 906 &1332 &427 & 8.77 & 8.86 &-21.50 &30.37 &29.93\\
 NGC3370                               &  32207 &1367 &1622 &311 &11.69 &11.81 &-20.26 &32.07 &32.09\\
 NGC3627                               &  34695 & 723 &1454 &384 & 8.26 & 8.36 &-21.08 &29.44 &29.69\\
 NGC3663                               &  35006 &5040 &5389 &443 &12.42 &12.37 &-21.65 &34.07 &34.24\\
 NGC3672                               &  35088 &1860 &2210 &399 &10.57 &10.66 &-21.23 &31.89 &32.20\\
 NGC4501                               &  41517 &2268 &1740 &570 & 8.75 & 8.85 &-22.64 &31.49 &30.93\\
 NGC4527                               &  41789 &1736 &2090 &361 & 9.32 & 9.56 &-20.84 &30.39 &30.42\\
 NGC4536                               &  41823 &1808 &2162 &341 & 9.81 & 9.95 &-20.61 &30.56 &30.75\\
 NGC4639                               &  42741 &1003 &1740 &348 &11.25 &11.26 &-20.69 &31.96 &31.80\\
 NGC4680                               &  43118 &2491 &2811 &237 &12.10 &12.24 &-19.17 &31.41 &32.54\\
 NGC4679                               &  43170 &4665 &3824 &426 &11.72 &11.84 &-21.49 &33.36 &33.89\\
 NGC5005                               &  45749 &1011 &1177 &601 & 9.01 & 9.08 &-22.85 &31.93 &31.17\\
 ESO576-040                          &  46574 &2095 &2407 &169 &13.72 &13.61 &-17.85 &31.47 &31.89\\
 PGC047514                           &  47514 &4217 &4577 &284 &13.96 &13.82 &-19.89 &33.75 &34.34\\
 NGC5584                        & 51344 &1655 & 191 &266 &11.74 &11.72 & -19.64 & 31.35 & 31.92 \\
 \textcolor{red}{IC4423}         &   \textcolor{red}{51549} & \textcolor{red}{9115} & \textcolor{red}{9691} & \textcolor{red}{470} & \textcolor{red}{13.73} & \textcolor{red}{13.92} & \textcolor{red}{-21.88} & \textcolor{red}{35.95} & \textcolor{red}{35.67}\\
 IC1151                      &  56537 &2176 &2287 &241 &12.83 &12.83 &-19.25 &32.08 &33.16\\
  NGC6063                              &  57205 &2841 &2958 &308 &12.98 &12.95 &-20.21 &33.18 &32.99\\
 UGC10738                             &  59769 &6716 &6850 &584 &12.37 &12.53 &-22.74 &35.38 &34.85\\
 UGC10743                             &  59782 &2744 &2581 &218 &12.59 &12.76 &-18.85 &31.61 &32.68\\
 NGC6962                               &  65375 &4200 &3695 &639 &11.05 &11.15 &-23.09 &34.31 &33.69\\
 IC5179                &  68455 &3400 &3108 &444 &10.80 &11.14 &-21.66 &32.81 &33.18\\
 \textcolor{red}{UGC12133}  &   \textcolor{red}{69428} & \textcolor{red}{7391} & \textcolor{red}{7213} & \textcolor{red}{442} & \textcolor{red}{13.17} & \textcolor{red}{13.32} & \textcolor{red}{-21.64} & \textcolor{red}{35.05} & \textcolor{red}{34.99}\\
NGC7329                                &  69453 &3245 &3150 &461 &11.19 &11.34 &-21.80 &33.16 &33.19\\
NGC7448                               &  70213 &2170 &1752 &309 &11.32 &11.40 &-20.23 &31.63 &32.72\\
\hline
\end{tabular}
\end{center}
\caption{Properties of individual SNIa galaxies (latest results): (1) Common name, (2) PGC name, (3) Mean velocity of host galaxy with respect to the CMB, \kms, (4) Mean velocity of host galaxy with respect to the CMB corrected for the cosmological model, \kms, (5) Corrected rotation rate parameter corresponding to twice the maximum velocity, \kms, (6) Corrected $3.6\ \mu$m magnitude in the AB system, mag, (7) Color adjusted magnitude, mag, (8) Absolute color adjusted magnitude, mag, (9) TFR distance modulus corrected for bias, mag, (10) SNIa distance modulus, mag. Supplementary galaxies with respect to the 2012 work are in red.}
\label{tblCh3:SNIav2}
\end{table*}

\begin{table*}
\begin{center} 
\begin{tabular}{|r|l|r|r|r|c|c|c|c|c|c|}
\hline
   PGC  &   Name& I$_{ext}^{b,i,k}$& [3.6]$_{ave}^{b,i,k,a}$& C$_{[3.6]_{ave}}^{b,i,k,a}$ & b/a& Inc &W$_{mx}$& W$_{mx}^i$ &W$_{mxl}^i$&  Sample\\
\hline
   2758   & NGC0247  &  7.79  &    9.10   & 8.98   &   0.31 &  76. &190 &196 &  2.292    & ZP \\
   3238   & NGC0300  &  7.28  &    8.40   & 8.38   &   0.71 &  46. &140 &195 &  2.290    & ZP \\
   9332   & NGC0925  &  8.96  &   10.25  & 10.14 &   0.57 &  57. &194 &231 &  2.364    & ZP \\
  13179  & NGC1365  &  8.09  &    8.77   & 8.97   &  0.61  & 54. & 371&459 &  2.662    & ZP \\ 
  13602  & NGC1425  &  9.50  &   10.72 &  10.64 &   0.46 &  65. & 354 &391 &  2.592    & ZP \\ 
  17819  & NGC2090  &  9.33  &   10.38 &  10.39 &   0.43 &  67. & 277 & 301&  2.478    & ZP \\ 
  21396  & NGC2403  &  7.11  &    8.46   &  8.32  &   0.53 &  60. &226 &261 &  2.417    & ZP \\ 
  23110  & NGC2541  & 10.76 &   12.06 & 11.94 &   0.49 &  63. & 188& 211&  2.325    & ZP \\ 
  26512  & NGC2841  &  7.53  &    8.65   & 8.63   & 0.45   &66. &592 &650 &  2.813    & ZP \\ 
  28120  & NGC2976  &  8.98  &    9.89   & 9.97   & 0.53   &60. &129 &149 &  2.173    & ZP \\ 
  28630  & NGC3031  &  5.20  &    6.29   & 6.28   & 0.54   &59. & 416&485 &  2.686    & ZP \\ 
  30197  & NGC3198  &  9.17  &   10.33 & 10.28 &   0.39 &  70. & 296&315 &  2.498    & ZP \\ 
  30819  & IC2574       &10.12 &    11.12 & 11.16 &   0.40 &  69. &106&113 &  2.054    & ZP \\ 
  31671  & NGC3319  & 10.55 &   11.82 & 11.72 &   0.54 &  59. &195 & 227 &  2.356    & ZP \\ 
  32007  & NGC3351  &  8.33  &    9.20   & 9.31   & 0.70   &47. & 262&312&  2.556    & ZP \\ 
  32192  & NGC3368  &  7.88  &    8.80   & 8.88   & 0.64   &52. & 329& 418 &  2.621    & ZP \\ 
  34554  & NGC3621  &  8.01  &    9.01   & 9.05   & 0.45   &66. & 266&292 &  2.465    & ZP \\ 
  34695  & NGC3627  &  7.39  &    8.28   & 8.38   & 0.53   &60. & 333&385 &  2.585    & ZP \\ 
  39422  & NGC4244  &  8.92  &   10.25 &  10.12 &   0.20 &  90. &192 & 192&  2.283    & ZP \\ 
  39600  & NGC4258  &  6.84  &    7.98   & 7.95   & 0.40   &69. &414 & 444&  2.647    & ZP \\ 
  40596  & NGC4395  &  9.08  &   11.21 & 10.66 &   0.73 &  44. &112 & 161&  2.206    & ZP \\ 
  40692  & NGC4414  &  8.73  &    9.38   & 9.60   & 0.60   &55. &378 &463&  2.666    & ZP \\ 
  41812  & NGC4535  &  8.95  &    9.75   & 9.89   & 0.72   &45. & 265& 374&  2.573    & ZP \\ 
  41823  & NGC4536  &  9.03  &    9.85   & 9.98   & 0.38   &71. &322 &341 &  2.533    & ZP \\ 
  42408  & NGC4605  &  9.19  &   10.17 & 10.22 &   0.41 &  69. & 154& 165&  2.219    & ZP \\ 
  42510  & NGC4603  &  9.76  &   10.67 & 10.75 &   0.64 &  52. &353 &450 &  2.653    & ZP \\ 
  42741  & NGC4639  & 10.18 &   11.27 & 11.26 &   0.60 &  55. & 274&336 &  2.526    & ZP \\ 
  43451  & NGC4725  &  7.84  &    8.87   &  8.89  &  0.56  & 58. & 397& 470&  2.672    & ZP \\ 
  47368  & NGC5204  &   /        &  11.93  &    /       &  0.50  & 62. &186 &267 &  2.095    & ZP \\ 
  60921  & NGC6503  &  8.67  &    9.78   & 9.76   & 0.32   &75. & 223&231 &  2.363    & ZP \\
  69327  & NGC7331  &  7.52  &    8.39   & 8.50   & 0.44   &66. & 501& 547&  2.738    & ZP \\
  73049  & NGC7793  &  8.25  &    9.27   & 9.30   & 0.62   &53. & 162& 202&  2.306    & ZP \\
  ... & & & & & & & & & &\\
\hline
\end{tabular}
\end{center}
\caption{Calibrator parameters for the Tully-Fisher relation (complete table online): (1) PGC number, (2) Common Name, (3) I band corrected magnitude, mag, (4) [3.6] averaged corrected magnitude, mag, (5) Pseudo [3.6] magnitude, mag, (6) Axial Ratio, (7) Inclination, degrees, (8) linewidth not corrected for inclination, \kms, (9) linewidth corrected for inclination, \kms, (10) Logarithm of the inclination corrected linewidth, (11) Sample (ZP Zeropoint Calibrators)}
\label{CalibTFv2}
\end{table*}

\clearpage

\begin{figure*}
\vspace{-3cm}
\centering
\includegraphics[scale=1]{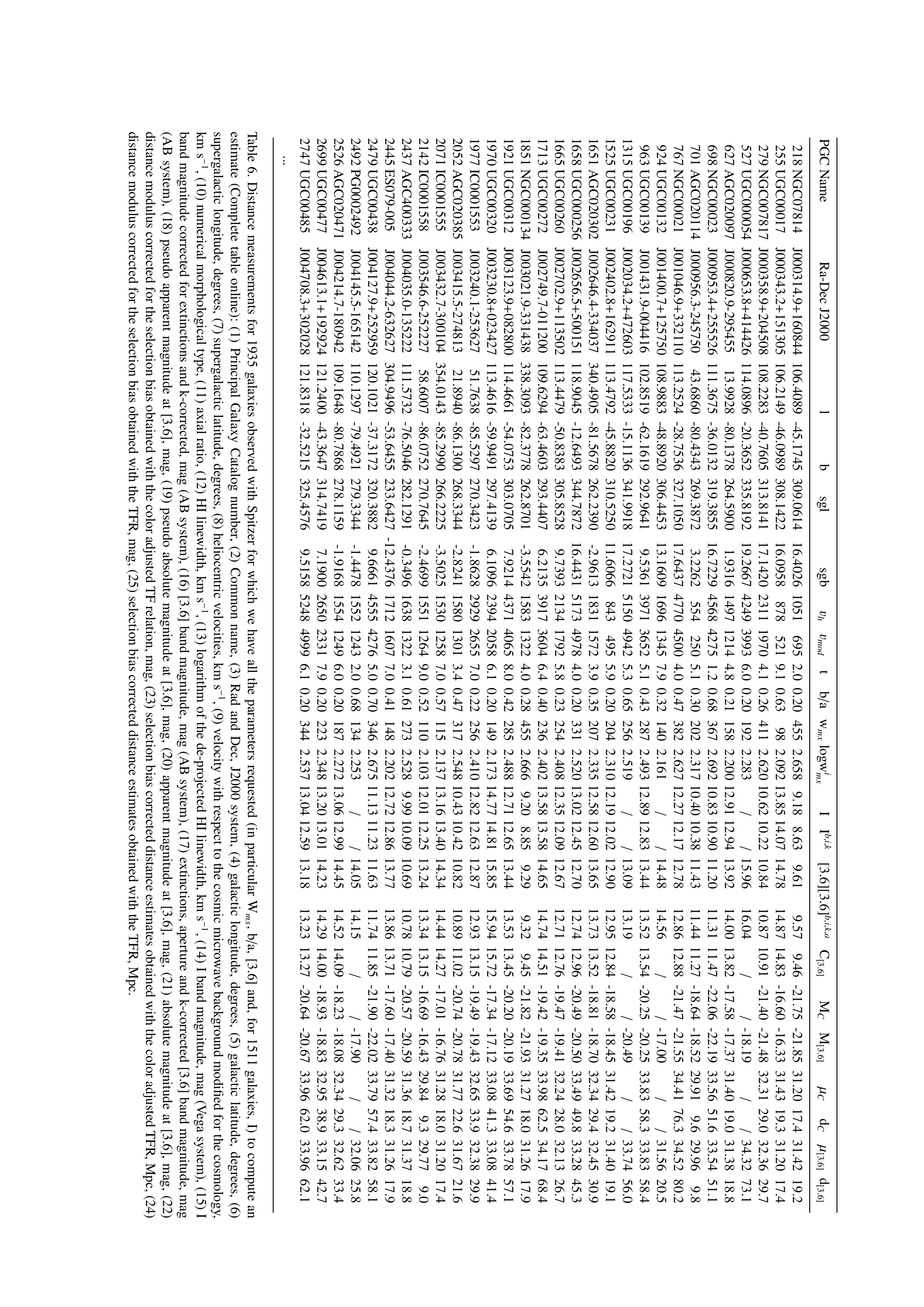}\\
\caption{}
\label{DistEstim}
\end{figure*}
\clearpage 
\newpage
\pagestyle{empty}
~
\newpage
\bibliographystyle{mn2e}

\bibliography{biblicomplete}

\end{document}